\PassOptionsToPackage{svgnames}{xcolor}
\documentclass[journal]{vgtc}                     % final (journal style)
\graphicspath{{figs/}{figures/}{pictures/}{images/}{./}} % where to search for the images

\usepackage{times}                     % we use Times as the main font
\usepackage{tabu}                      % only used for the table example
\usepackage{booktabs}                  % only used for the table example
\usepackage{lipsum}                    % used to generate placeholder text
\usepackage{mwe}                       % used to generate placeholder figures
\usepackage{pifont}% http://ctan.org/pkg/pifont
\usepackage{multirow}
\usepackage[most]{tcolorbox}
\usepackage{enumitem}
\usepackage{tcolorbox}
\usepackage{subcaption}
\usepackage{stfloats}
\usepackage{mathptmx}                  % use matching math font

\onlineid{1411}

\vgtccategory{Research}

\vgtcinsertpkg

\title{\corpus: An SVG Chart Corpus with\\Fine-Grained Semantic Labels}

\author{
Chen Chen, Hannah K. Bako, Peihong Yu, John Hooker, Jeffrey Joyal, Simon C. Wang, Samuel Kim, \\Jessica Wu, Aoxue Ding, Lara Sandeep, Alex Chen, Chayanika Sinha, Zhicheng Liu%
\thanks{All authors are with the Department of Computer Science, University of Maryland, College Park, MD, USA. Emails: \{cchen24, hbako, peihong\}@umd.edu, \{jhooker, jjoyal, scwang00, skim1270, jwu36, ading1, lsandeep, achen131, csinha\}@terpmail.umd.edu, \{leozcliu\}@umd.edu}
}

\teaser{
  \centering\includegraphics[width=\linewidth]{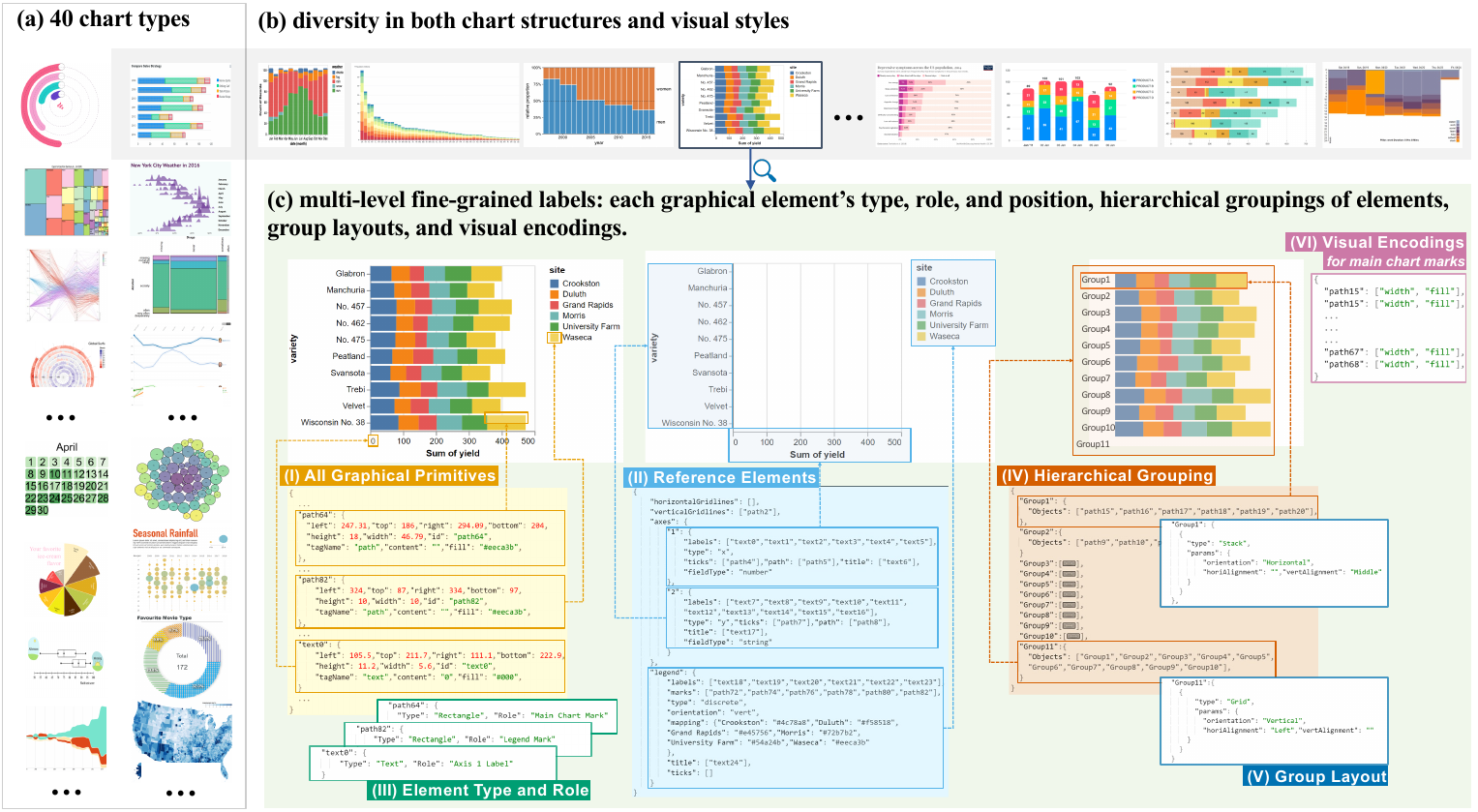}
  \vspace{-5mm}
  \caption{\corpus{} is an SVG chart corpus that (a) consists of $942$ charts across $40$ chart types, (b) promotes diversity within each chart type, featuring rich design variations in terms of visual structures and styles, and (c)  distinguishes from existing chart corpora in its multi-granular chart semantic labels on more than 383K graphical elements, including each element's type, role, and position, hierarchical groupings of elements, group layouts, and visual encodings.}
  \label{fig:teaser}
}

\abstract{
    Chart corpora, which comprise data visualizations and their semantic labels, are crucial for advancing visualization research.
However, the labels in most existing corpora are high-level (e.g., chart types), hindering their utility for broader applications in the era of AI. 
In this paper, we contribute \corpus, a corpus containing $942$ real-world SVG charts produced by over $50$ tools, encompassing $40$ chart types and featuring structural and stylistic design variations. 
Each chart is augmented with multi-level fine-grained labels on its semantic components, including each graphical element's type, role, and position, hierarchical groupings of elements, group layouts, and visual encodings. In total, \corpus provides labels for more than $383$k graphical elements.
We demonstrate the richness of the semantic labels by comparing \corpus with existing corpora. 
We illustrate its usefulness through four applications: semantic role inference for SVG elements, chart semantic decomposition, chart type classification, and content navigation for accessibility. 
Finally, we discuss research opportunities to further improve \corpus.
} % end of abstract


\usepackage{xargs}      % Used for new commands with optional arguments
\usepackage{soul}       % Used for custom comments
\usepackage{color}      % Used for custom colors in comments
\usepackage{xspace}     % Used for abbreviation spacing
\usepackage{xpunctuate} % Used for abbreviation spacing
\usepackage{circledsteps}  % Used for creating document figure labels

\newcommand{\corpus}{\textcolor{black}{\mbox{\textsc{VisAnatomy}}}\xspace}

\newcommand{\cmark}{\ding{51}}

\definecolor{stepLabelGreen}{RGB}{0,158,115}
\newcommand{\exist}{\Circled[inner color=white, fill color=stepLabelGreen, outer color=stepLabelGreen]{\cmark}}
\newcommand{\partialExist}{\Circled[inner color=white, fill color=orange, outer color=orange]{\cmark}}
\newcommand{\canExist}{\Circled[inner color=white, fill color=gray, outer color=gray]{\cmark}}

\definecolor{boxOrange}{RGB}{230,159,0}
\definecolor{boxSkyBlue}{RGB}{86, 180, 233}
\definecolor{boxBluishGreen}{RGB}{0, 158, 115}
\definecolor{boxVermilion}{RGB}{213, 94, 0}
\definecolor{boxBlue}{RGB}{0, 114, 178}
\definecolor{boxReddishPurple}{RGB}{204, 121, 167}

\newcommand{\supp}{supplementary materials}
\newcommand{\stageAL}{\textsf{\small \textbf{Axis\&Legend}}}
\newcommand{\stageM}{\textsf{\small \textbf{Marks}}}
\newcommand{\stageG}{\textsf{\small \textbf{Groups}}}
\newcommand{\stageL}{\textsf{\small \textbf{Layout}}}
\newcommand{\stageE}{\textsf{\small \textbf{Encodings}}}

\NewDocumentCommand{\allElement}{m}{%
  \tcbox[
    colback=boxOrange!75,
    colframe=boxOrange!75,
    boxrule=0.5pt,
    arc=0.5pt,
    boxsep=0pt,
    left=0.25pt,
    right=0.25pt,
    top=0.25pt,
    bottom=0.25pt,
    tcbox raise base,
    nobeforeafter,
    height=1em,
    valign=center
  ]{\color{white} \small All Graphic Primitives}%
}
\newcommand{\allEle}{\allElement{}}

\NewDocumentCommand{\referElement}{m}{%
  \tcbox[
    colback=boxSkyBlue!66,
      colframe=boxSkyBlue!66,
      boxrule=0.5pt,
    arc=0.5pt,
    boxsep=1.45pt,
    left=-1.25pt,
    right=-1.25pt,
    top=0.25pt,
    bottom=0.25pt,
    tcbox raise base,
    nobeforeafter,
    height=1em,
    valign=center
  ]{\color{white} \small Reference Elements}%
}
\newcommand{\refEle}{\referElement{}}

\NewDocumentCommand{\markInformation}{m}{%
  \tcbox[
    colback=boxBluishGreen!66,
      colframe=boxBluishGreen!66,
      boxrule=0.5pt,
    arc=0.5pt,
    boxsep=0pt,
    left=0.25pt,
    right=0.25pt,
    top=0.25pt,
    bottom=0.25pt,
    tcbox raise base,
    nobeforeafter,
    height=1em,
    valign=center
  ]{\color{white} \small Element Type and Role}%
}
\newcommand{\markInfo}{\markInformation{}}

\NewDocumentCommand{\elementTypes}{m}{%
  \tcbox[
    colback=boxBluishGreen!66,
      colframe=boxBluishGreen!66,
      boxrule=0.5pt,
    arc=0.5pt,
    boxsep=0pt,
    left=0.25pt,
    right=0.25pt,
    top=0.25pt,
    bottom=0.25pt,
    tcbox raise base,
    nobeforeafter,
    height=1em,
    valign=center
  ]{\color{white} \small Element Type}%
}
\newcommand{\markType}{\elementTypes{}}

\NewDocumentCommand{\elementRoles}{m}{%
  \tcbox[
    colback=boxBluishGreen!66,
      colframe=boxBluishGreen!66,
      boxrule=0.5pt,
    arc=0.5pt,
    boxsep=1.45pt,
    left=-1.25pt,
    right=-1.25pt,
    top=0.25pt,
    bottom=0.25pt,
    tcbox raise base,
    nobeforeafter,
    height=1em,
    valign=center
  ]{\color{white} \small Element Role}%
}
\newcommand{\markRole}{\elementRoles{}}

\NewDocumentCommand{\groupInformation}{m}{%
  \tcbox[
    colback=boxVermilion!66,
      colframe=boxVermilion!66,
      boxrule=0.5pt,
    arc=0.5pt,
    boxsep=0pt,
    left=0.25pt,
    right=0.25pt,
    top=0.25pt,
    bottom=0.25pt,
    tcbox raise base,
    nobeforeafter,
    height=1em,
    valign=center
  ]{\color{white} \small Hierarchical Grouping}%
}
\newcommand{\groupInfo}{\groupInformation{}}

\NewDocumentCommand{\layoutInformation}{m}{%
  \tcbox[
    colback=boxBlue!66,
      colframe=boxBlue!66,
      boxrule=0.5pt,
    arc=0.5pt,
    boxsep=0pt,
    left=0.25pt,
    right=0.25pt,
    top=0.25pt,
    bottom=0.25pt,
    tcbox raise base,
    nobeforeafter,
    height=1em,
    valign=center
  ]{\color{white} \small Group Layout}%
}
\newcommand{\layoutInfo}{\layoutInformation{}}

\NewDocumentCommand{\encodingInformation}{m}{%
  \tcbox[
    colback=boxReddishPurple!66,
      colframe=boxReddishPurple!66,
      boxrule=0.5pt,
    arc=0.5pt,
    boxsep=0pt,
    left=0.25pt,
    right=0.25pt,
    top=0.25pt,
    bottom=0.25pt,
    tcbox raise base,
    nobeforeafter,
    height=1em,
    valign=center
  ]{\color{white} \small Visual Encodings}%
}
\newcommand{\encInfo}{\encodingInformation{}}

\newcommand{\ie}{{i.e.,}\xspace}
\newcommand{\eg}{{e.g.,}\xspace}
\newcommand{\etal}{{et~al\xperiod}\xspace}

\newcommand{\etc}{{etc\xperiod}\xspace}

\newcommand{\parHeading}[1]{\vspace{2px}\noindent{\textbf{#1}}}

\definecolor{lightpink}{RGB}{237,157,202}
\definecolor{lightred}{RGB}{210,121,121}
\definecolor{lightorange}{RGB}{230,170,50}
\definecolor{lightgold}{RGB}{210,194,121}
\definecolor{lightgreen}{RGB}{121,210,121}
\definecolor{lightaqua}{RGB}{121,206,210}
\definecolor{lightBlue}{RGB}{121,124,210}
\definecolor{lightpurple}{RGB}{153,102,255}
\definecolor{red}{RGB}{178,34,34}
\definecolor{gray}{RGB}{166,166,166}

\newcommand{\gone}{\textsf{\small Graph 1}}
\newcommand{\gtwo}{\textsf{\small Graph 2}}
\newcommand{\gthree}{\textsf{\small Graph 3}}
\newcommand{\gfour}{\textsf{\small Graph 4}}

\newcommandx{\chen}[2][1=] 
    {\setulcolor{lightgreen}{\ul{#1}} \textcolor{lightgreen}   %% Usage: \jane[(optionally) underline text]{With a comment.}
    {[\textbf{Chen:} #2]}}

\begin{document}

\newcommand{\revision}[1]{\textcolor{black}{#1}}
\newcommand{\leo}[1]{\textcolor{olive}{Leo: #1}}
\newcommand{\hannah}[1]{\textcolor{teal}{Hannah: #1}}
%%%%%%%%%%%%%%%%%%%%%%%%%%%%%%%%%%%%%%%%%%%%%%%%%%%%%%%%%%%%%%%%
%%%%%%%%%%%%%%%%%%%%%% START OF THE PAPER %%%%%%%%%%%%%%%%%%%%%%
%%%%%%%%%%%%%%%%%%%%%%%%%%%%%%%%%%%%%%%%%%%%%%%%%%%%%%%%%%%%%%%%

% \firstsection{Introduction}

% \input{sections/01-Intro}
% \input{sections/02-overview}
% \input{sections/03-construction}
% \input{sections/04-use cases}
% \input{sections/05-discussion}
% \input{sections/06-conclusion}

\firstsection{Introduction}\label{sec:introV2}
\maketitle

Visualization researchers have been curating chart corpora to advance the state of the art in chart creation and generation~\cite{dibia2019data2vis,cui2019text}, classification~\cite{jung_chartsense_2017,savva_revision_2011}, retrieval~\cite{structure2022Li,hu2019vizml}, decomposition~\cite{mystique,poco_reverse-engineering_2017}, and editing~\cite{mystique,cui_mixed-initiative_2022}. The availability of fine-grained semantic labels such as element properties and data encodings is vital for a 
% The absence of fine-grained semantics in visualization labels constrains the 
range of visualization downstream tasks.
% , especially when the tasks involve multi-level semantics and human interactions. 
For example, the shapes and roles of visual elements as well as their visual properties are required to \textit{develop (semi-)automated data visualization reuse pipelines}~\cite{mystique, savva_revision_2011}; the correspondence between visual elements~(or groups) and axis/legend labels is necessary for \textit{developing chart reader experiences for visually impaired people}~\cite{screenReader}; the grouping structure of visual elements can be utilized to develop graph neural network models~\cite{structure2022Li}. 

However, the semantic labels in existing chart corpora are often insufficient and sometimes unreliable for supporting various visualization tasks \cite{mystique,chen2023state}.
According to two recent surveys~\cite{chen2023state,davila2020chart}, 
many corpora are not publicly available. The remaining ones typically have only high-level labels (e.g., chart types~\cite{battle2018beagle,savva_revision_2011} and chart area bounding boxes~\cite{deng_visimages_2022}).  
Although a few existing corpora do offer fine-grained labels, they often exhibit limited diversity in terms of the variety of chart-authoring tools and chart designs. 
% For instance, the YOLaT++ chart corpus contains charts produced by only two tools \cite{dou2024hierarchical}. 
% the corpus used in Chartreuse~\cite{cui_mixed-initiative_2022} 
% and Li~\etal \cite{structure2022Li} encompass 
% contains infographic bar charts exclusively from Microsoft PowerPoint.
% and Plotly~\cite{Plotly}, respectively.  
This limited diversity hinders the generalizability of models built upon those corpora: 
they can easily fail when handling ``out-of-distribution (OOD)'' charts  produced by other tools with inconsistent usage of SVG elements~\cite{structure2022Li} and grouping structures~\cite{mystique}.
In general, existing chart corpora are inadequate to support the development of robust visualization applications.
% that use SVG charts as inputs.

In this paper, we seek to address these limitations and contribute a diverse SVG chart corpus, \corpus, with multi-level fine-grained semantic labels. 
\corpus includes 942 real-world SVG charts
% ~(also available in the raster image format) 
and their corresponding multi-level semantic labels. 
The underlying data tables are also included if available (329 out of 942). 
The charts are collected through a manual process with careful inspection. 
For each chart, multiple independent expert annotators use a semi-automated annotation tool to obtain the semantic labels, and the quality of the labels is controlled through consensus among them. 

\corpus makes two key extensions over prior chart corpora.
First, regarding corpus diversity, \corpus encompasses 40 chart types~(synthesized from three visualization typologies~\cite{chartmaker_dir,dataVizCata,dataVizProj}) produced by over 50 tools from hundreds of public online sources, featuring structural and stylistic design variations~(Section~\ref{sec:collecting}).
Second, and more importantly, we identified a set of core components through a survey on existing visualization scene models~\cite{liu2024msc,satyanarayan_critical_2019,satyanarayan_vega-lite_2016,snyder_divi_2024} to augment each chart in \corpus with comprehensive semantic labels, including each visual element's shape (e.g., rectangle, pie, polyline), role (e.g., main chart mark, axis path, legend tick, annotation), and bounding box, the hierarchical grouping of elements, the layout for each group, and visual encodings~(Section~\ref{sec:deisredlabels}). In total, \corpus provides labels for $383,459$ visual elements.

We evaluate the richness of semantic labels and corpus diversity through a comparison between \corpus and existing corpora~(Section~\ref{sec:comparison}). 
Four applications illustrate the usefulness of \corpus\ (Section~\ref{sec:useCases}): \revision{semantic role inference for SVG elements}, chart semantic decomposition, chart type classification, and content navigation for accessibility. 
Finally, we discuss the current limitations of \corpus\ and outline our future work
% including enlarging the size of \corpus, obtaining labels of relational constraints, extracting data tables, and exploring visualization applications in greater depth and breadth~
(Section~\ref{sec:discussion}). The \corpus corpus is available at \href{https://VisAnatomy.github.io/}{\color{blue!80!black}{https://VisAnatomy.github.io/}}. 
\section{\corpus: Chart Collection}\label{sec:collecting}

In this section, we introduce the process to construct \corpus. 
We first describe how the team decided on the desired chart types and collected charts following a standardized procedure.
We then give an overview of the 942 charts in \corpus, showing the distributions of the chart types, charting tools, and source domains.

\subsection{Manual Chart Collection}\label{sec:chartCollection} 
% \leo{change all verbs to past tense}
Before the collection process started, we decided to focus on charts in the SVG (Scalable Vector Graphics) format. In recent years, SVG has emerged as a popular choice for curating corpora with fine-grained semantic labels in diverse visualization applications \cite{mystique,structure2022Li,masson2023chartdetective,poco_reverse-engineering_2017,snyder_divi_2024}. Compared to raster images, SVG includes low-level details such as element types and visual styles in its XML structure~\cite{chen2023state}, removing the need for error-prone image segmentation~\cite{Minaee2022segmentation} and element extraction~\cite{ocr}. Compared to code, where a label extraction approach is not easily generalizable to different visualization languages~\cite{andrew2023no}, SVG is supported as the output format by a wide array of languages and tools.

We manually searched for and sampled real-world SVG charts online to form the chart collection in~\corpus. 
There are other approaches to collecting charts, such as web crawling, that could lead to larger corpora. 
However, charts collected through automatic crawling cannot guarantee
a balanced distribution of charts or diversity in terms of chart types and design variations~\cite{chen2023state}. 
Also, since SVGs can be embedded in web pages in various ways (\eg~inline SVG, object tag, iframes), consistent extraction of SVG charts in practice is more challenging compared to raster images, which are mostly directly embedded. 
% \leo{also explain that crawling works better for images, less so for SVGs (explain why)} 
We therefore decided to follow a manual collection process to ensure the quality and diversity of the SVG charts.

The first step in our chart collection process was to compile a set of targeted chart types by browsing three visualization typologies: the Chartmaker Directory~\cite{chartmaker_dir}, the Data Viz Project~\cite{dataVizProj}, and the Data Visualisation Catalogue~\cite{dataVizCata}, each of which contains a detailed categorization of chart types. We cross-checked the named chart types in these three collections and focused on visualizations that consist of basic geometric shapes including \texttt{rectangle}, \texttt{circle}, \texttt{ellipse}, \texttt{pie}, \texttt{arc}, \texttt{line}, \texttt{polyline}, \texttt{area}, \texttt{polygon}, \texttt{geo-polygon}, and \texttt{text}. 
For the underlying data in the visualizations, Munzner~\cite{munzner_visualization_2014} defined four types of datasets: tables (items \& attributes), trees and networks (nodes \& links), fields, and geometry. We decided to focus on visualizations of tables, thus excluding node-link diagrams and scientific visualizations. 
% We exclude 3D geometries as scientific visualizations lead to very different design variations and are beyond the scope of this work. 
% We also exclude node-link diagrams, such as Sankey diagrams and tree diagrams, because their algorithmic layouts are deeply coupled with the underlying data formats/values and require a different semantic label set to describe compared to other chart types. \leo{we may need a stronger reason} \hannah{agree.} \chen{not sure if it is now stronger? found it hard to justify this point}
% \leo{Describe how many chart types from each of the typologies meet our criteria} 
Applying these criteria, we narrowed down to 31, 37, and 37 chart types from the three typologies, respectively.
Finally, we unified the names of these chart types and achieved a final set of 40 chart types~(Figure~\ref{fig:chartTypeDist}).

We then started collecting SVG charts for each chart type by (1) browsing online charting tool galleries~(e.g., D3.js~\cite{bostock_d3_2011}, Vega-Lite~\cite{satyanarayan_vega-lite_2016}, Mascot.js~\cite{liu_atlas_2021}), (2) browsing online communities where visualizations are shared by chart makers~(e.g., Observable~\cite{observable}, bl.ocks.org~\cite{blockOrg}, Spotfire~\cite{spotfire}), and (3) searching for certain chart types using engines such as Google Advanced Image Search~\cite{GoogleImage}~(with ``SVG'' specified as the target file type) and Bing Visual Search~\cite{Bing}.
% \leo{say more about the manual search and curation process, e.g., charting tool galleries, Google search with SVG as target file format, etc. Refer to meeting notes.} 
It is important to note that when collecting charts, \textit{we prioritized diversity over quantity because promoting rich design variations and supporting broader visualization applications are our research goals}. 
Therefore, for each chart type, we focused on including designs from different visualization galleries and websites, showcasing diverse visual styles.
We aimed to include at least 20 designs for each chart type. However, we kept adding new designs as long as they introduced unique visual styles that had not yet been observed and did not distort the overall distribution of chart types~(\ie an approximately uniform distribution). 
% \fy{was there a target distribution? like approximately uniform?}
We also examined the details of SVG files to discard invalid ones, \eg those using \texttt{<image>} elements with hyperlinks to render the whole chart.
\revision{When the appearance of an SVG is influenced by webpage style rules (i.e., CSS), we would first attempt to manually inject the relevant styles into the SVG. If that was not feasible, the SVG was discarded. Note that we mainly focused on styles vital for chart integrity (e.g., color, stroke), and did not handle all the font-related styles in this process.}
% \revision{Describe how the global styles in a webpage are handled. We may not have handled the font related styles}
% , which does not carry sufficient semantics information. 
For each SVG chart collected, we obtained its image in the PNG format with a Python script.

The project team consisting of six authors held weekly meetings to inspect the collected SVG charts, removed unqualified charts~(e.g., repeated or highly similar charts), and enforced consistent selection criteria. 
% To ensure consistency, each type is collected by one member of the project team, and w
Charts that do not fall into one of the 40 types but still embody interesting design ideas were moved to the ``Others'' category where we store bespoke chart designs.
This process was repeated until we had the expected number of valid designs (\ie20) for each chart type.
% Charts that were discarded during the meetings would be replaced by new ones 
% found according the same criterion by the same collector 
% and inspected by the team again. 
% \hannah{I think we need to distinguish between ``charts'' (i.e., individual designs) and chart type. Maybe we could use design for the first instance to avoid confusion.}
This iterative process ensures that the whole team has at least one pass on every collected chart and that issues with collected charts are resolved consistently.
After a 28-week chart collection effort, we reached the final set of charts in both the vector and raster image formats in \corpus, containing a total number of $942$ charts with an approximately even distribution across the 40 chart types plus an ``Others'' category, encompassing more than $50$ charting tools and hundreds of web domains~(chart sources). 

% \hannah{I think this flow is okay, but I wonder if introducing the chart collection process first and then discussing what the overall collection looked like after would be better? } \leo{Maybe talk about the process first.}

\subsection{\corpus Promotes Chart Diversity}\label{sec:diversity}
\begin{figure}[ht]
\centering
\includegraphics[width=0.465\textwidth]{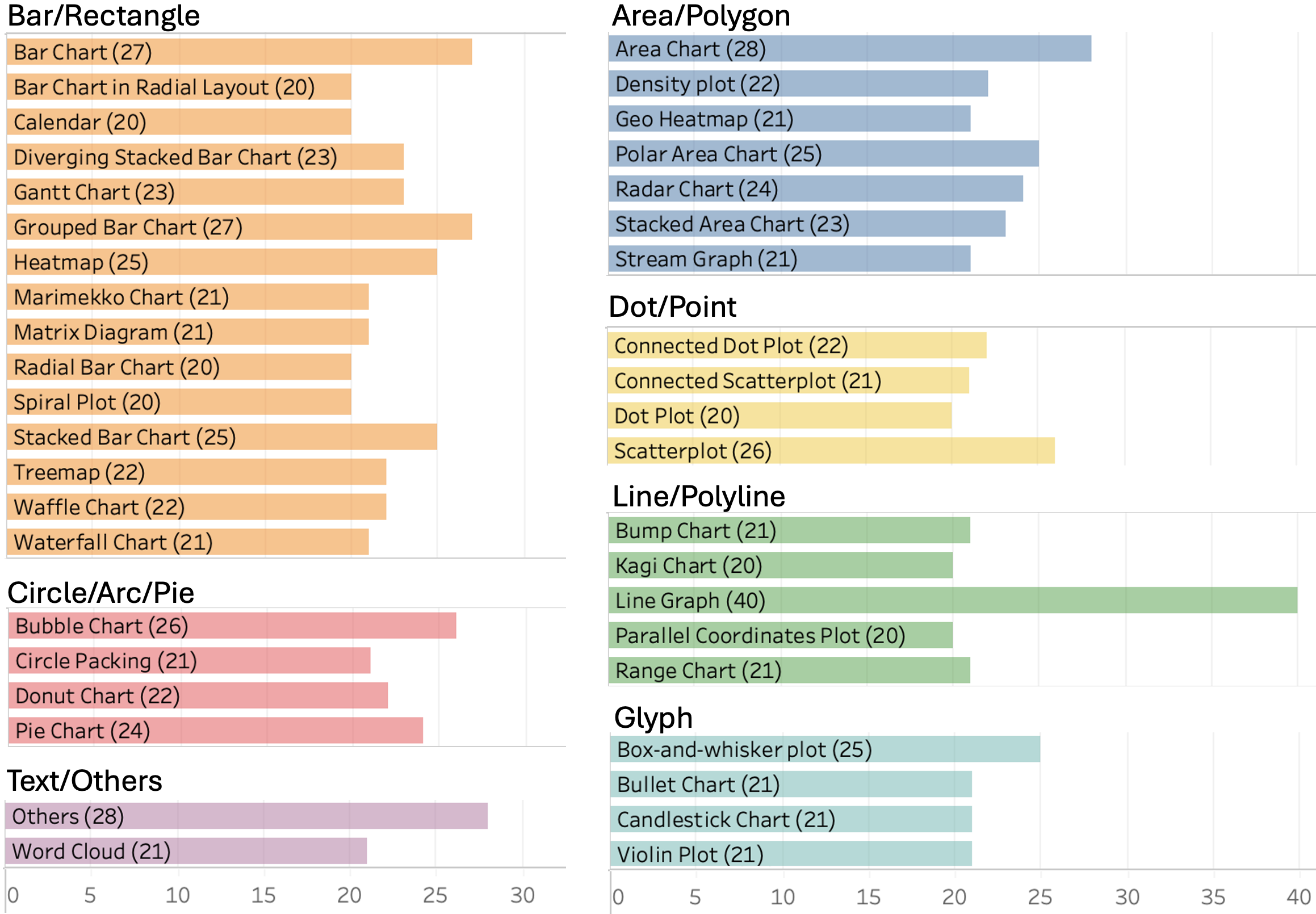}
\caption{Chart type distribution in \corpus categorized by primary mark types. Within each category the chart types are sorted alphabetically.}
\label{fig:chartTypeDist}
\end{figure}

\parHeading{Chart Types.} In its current state, \corpus\ contains 942 real-world SVG charts. Each chart has its corresponding label file in the JSON format, and is also available in the PNG (Portable Network Graphics) format. Figure \ref{fig:chartTypeDist} shows the overall distribution of the chart types categorized by the mark types. Within \corpus, there are $40$ named chart types together with an ``Others'' category containing custom designs such as composite or superimposed charts that do not fall appropriately into a specific chart type. 
Line graph, area chart, bar chart, grouped bar chart, and the "Others" category constitute higher proportions in the corpus, and the number of charts in each remaining type ranges between $20$ to $26$.
This distribution makes \corpus a balanced corpus with a wide variety of chart types.

\vspace{-5mm}

\begin{figure}[ht]
\centering
\includegraphics[width=0.49\textwidth]{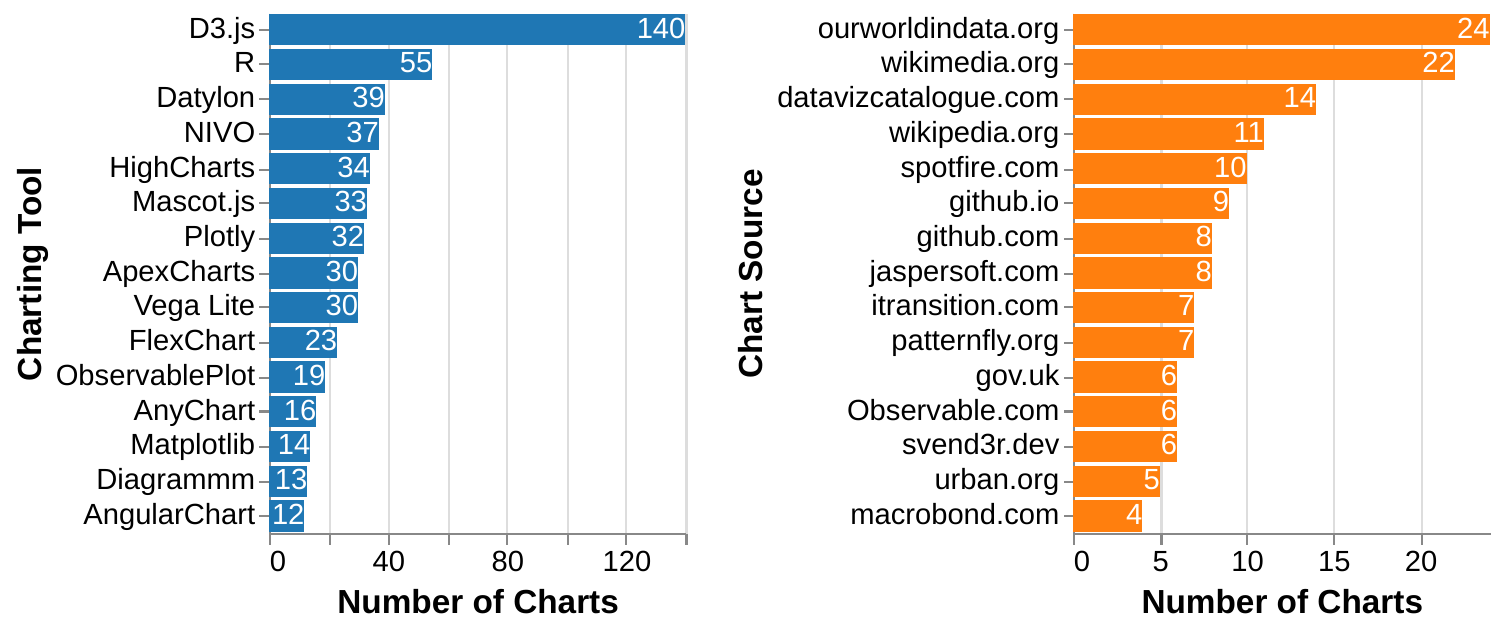}
\caption{Distributions of chart types, charting tools~(top-15), and source domains~(top-15) in \corpus.}
\vspace{-3mm}
\label{fig:chartingToolDist}
\end{figure}

\parHeading{Charting Tools and Chart Sources.} 
% Charts in \corpus\ are created using a variety of charting tools and come from diverse public online sources. 
For each chart, we record information on the charting tool~(if it is explicitly revealed in the source website) and/or the website domain~(if it is not coming from a charting tool's gallery). In total, \corpus\ collects charts created using 58 charting tools and more than 100 different online domains. In Figure \ref{fig:chartingToolDist} we show the top 15 charting tools and chart sources. 
% \hannah{are there cases where the chart tool was unknown?} 
% It is within the expectation that
D3.js~\cite{bostock_d3_2011} contributes the largest portion as it is the most expressive tool for creating interactive web-based SVG visualizations~\cite{battle2018beagle}. Several other visualization grammars and tools also provide a decent amount of charts to \corpus, including R~\cite{wilkinson2011ggplot2}, Datylon~\cite{datylon}, NIVO~\cite{nivo}, Highcharts~\cite{highcharts}, Mascot.js~\cite{liu_atlas_2021}, Plotly~\cite{Plotly}, Apexcharts.js~\cite{apexcharts}, and Vega-Lite~\cite{satyanarayan_vega-lite_2016}. The number of charts is more evenly distributed across chart sources, as most domains occur fewer than five times.

\parHeading{Design Variations.} In \corpus, we also strive for chart design diversity within each chart type, featuring rich design variations in terms of both chart structures and visual styles. Figure \ref{fig:teaser}(b) shows nine exemplary design variations for the stacked bar chart type. In \corpus, design variations include but are not limited to: 

\begin{itemize}[itemsep=0pt, topsep=0pt, parsep=0pt]
    \item different mark types used to create charts of the same type (e.g., bump charts composed of area marks or polylines with dots);
    \item different orientations of marks (e.g., connected dot plots containing horizontal or vertical glyphs);
    \item layering of marks (e.g., superimposed area charts);
    \item nested structures (e.g., small-multiple waffle charts and grouped box and whisker plots);
    \item different positions, orientations, and visual styles of reference elements such as axes, legends, and gridlines;
    \item different styles of annotations and embellishments.
\end{itemize}

\noindent We include detailed information of the chart types, tools, and sources in \corpus\ in the \supp{}\footnote{see \href{https://osf.io/962xc/?view_only=adbb315fd8794f6dac6b9625d385900f}{\color{blue}{detailed information}} of \corpus's chart collection.}.
% \leo{is sec 2.3 doing this now?}
\section{\corpus: Multilevel Fine-grained Semantic Labels}
\label{sec:labeling}

We took an iterative in-house labeling approach to obtain high-quality semantic labels for a set of core components synthesized from literature \cite{liu2024msc,satyanarayan_critical_2019,satyanarayan_vega-lite_2016,snyder_divi_2024}: mark elements reference elements, hierarchical grouping, visual encodings, and group layout. 
% based on the object model outlined in the Manipulable Semantic Components framework \cite{liu2024msc}. 
We decided not to obtain labels through crowdsourcing to ensure the label quality in~\corpus. Crowdsourced labels often require significant time to inspect and correct~\cite{crowdsourcing2008Kittur,kazai2013analysis,eventSeq2023Zinat}. Moreover, our desired labels require visualization expertise from the annotators, which is hard to guarantee through crowdsourcing platforms. Three experts in the team, each with at least four years of visualization research experience, participated in labeling the semantic components of the collected SVG charts. In this section, we describe the semantic labels that are associated with the charts in \corpus and the labeling process.

\subsection{Fine-grained Labels of Multilevel Scene Components}\label{sec:deisredlabels}
To support a broad set of visualization applications, we need detailed semantic labels beyond just chart types~\cite{chen2023state}. 
Specifically, we need a comprehensive understanding of the structure of a chart at multiple levels of granularity, from its global-level chart type down to the properties of individual elements. To this end, we surveyed related literature to compare existing visualization abstractions~\cite{liu2024msc,satyanarayan_critical_2019,satyanarayan_vega-lite_2016,snyder_divi_2024}.
% To be specific, a desired label of a chart would serve as a foundation, on which a global-to-local understanding of the chart representation can be built. 
% \begin{wrapfigure}{r}{0.5\textwidth}
%   \includegraphics[width=\linewidth]{figs/VOMlabel_v2.pdf}
%   \caption{The correspondence between our semantic labels and the VOM representation~\cite{liu_atlas_2021} of the stacked bar chart from Figure~\ref{fig:exampleLabel}. \leo{It might be better to show the chart again here side by side with the tree since the teaser is on page 1}}
%   \label{fig:vomlabel}
% \end{wrapfigure}
We finally decided to focus on the labels for the following components commonly shared across visualization grammars and abstractions: mark elements, reference elements, hierarchical grouping, visual encodings, and group layout. The supplemental materials contain detailed descriptions of the components we have surveyed.
% adopt the Visualization Object Model~(VOM) in the Manipulable Semantic Component~(MSC) framework proposed by Liu \etal~\cite{liu2024msc}, a graphics-centric approach to describing the structure of a visualization scene, because of (1) the more comprehensive component model it has compared to other candidates including D3.js~\cite{bostock_d3_2011} and Vega~\cite{vega_grammar}~(Table 1 in~\cite{liu2024msc}), and (2) its sufficient expressiveness (which is at the same level with Vega-Lite~\cite{satyanarayan_vega-lite_2016}).
% , as the theoretical framework to ground our labeling effort. 

% In MSC, the VOM consists of semantic components such as visual elements (\eg mark, glyph, collection), encodings (\ie mapping between data attribute and visual channel), algorithmic layouts (\eg grid, stack, packing, \etc) and relational constraints (\eg spatial alignment)~\cite{liu2024msc}.
% \leo{Provide a brief overview of the major scene components in the MSC framework.}

\begin{figure}[ht]
\centering
% \subfloat[\footnotesize The example stacked bar chart from Figure~\ref{fig:teaser}.\label{fig:teaserChart}]{
%     \includegraphics[width=0.455\textwidth]{figs/teaserChart.pdf}
% }
% \hfill
% \vspace{2mm}
% \subfloat[\footnotesize The correspondence between our semantic labels and the VOM representation~\cite{liu2024msc} of the example stacked bar chart. \label{fig:vomlabel}]{
\includegraphics[width=0.49\textwidth]{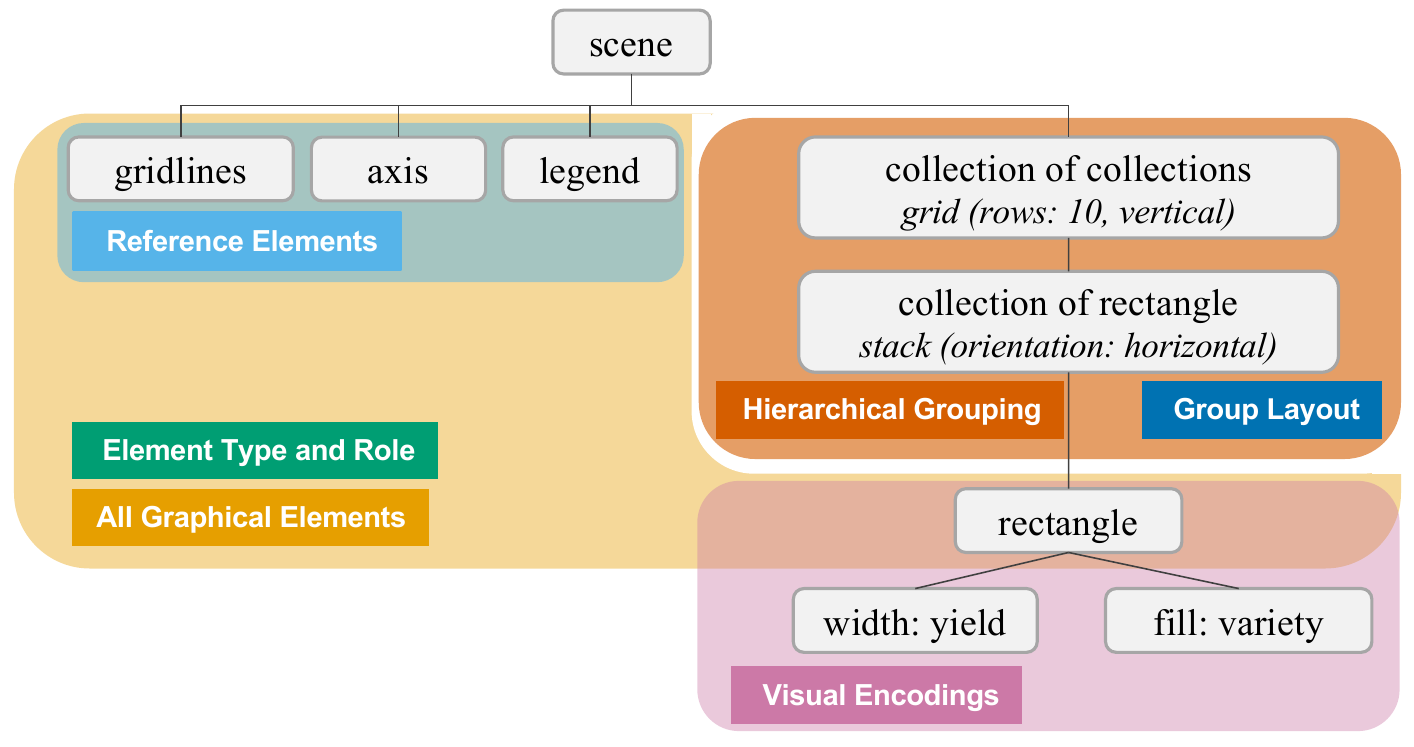}
% }
\caption{The components  in the stacked bar chart from Figure~\ref{fig:teaser}(c) and its correspondences to the labels in \corpus.}
\label{fig:chartVOMLabel}
\end{figure}

Using the stacked bar chart presented in Figure~\ref{fig:teaser}(c) as an example, we show its scene structure (gray nodes and edges) together with the correspondence to the semantic labels recorded in \corpus in Figure~\ref{fig:chartVOMLabel}.
Specifically, \refEle\ specify properties and involved elements for \texttt{gridlines}, \texttt{axes}, and \texttt{legend};
\allEle\ and \markInfo\ contain detailed information about all graphical elements in the \texttt{scene}, including \texttt{rectangle} marks and \refEle; \groupInfo\ and \layoutInfo\ correspond to the spatial clusters and relationships along the rightmost branch~(the \texttt{collection} subtree) of the scene graph, and \encInfo\ record encoded channels for \texttt{rectangle} marks. 
Note that in other types of visualizations, such correspondences remain similar. 
We next give a detailed explanation of the six semantic labels by walking through the example stacked bar chart~(in Figure~\ref{fig:teaser}(c)).

\begin{figure*}[!ht]
  \centering
  \includegraphics[width=\textwidth]{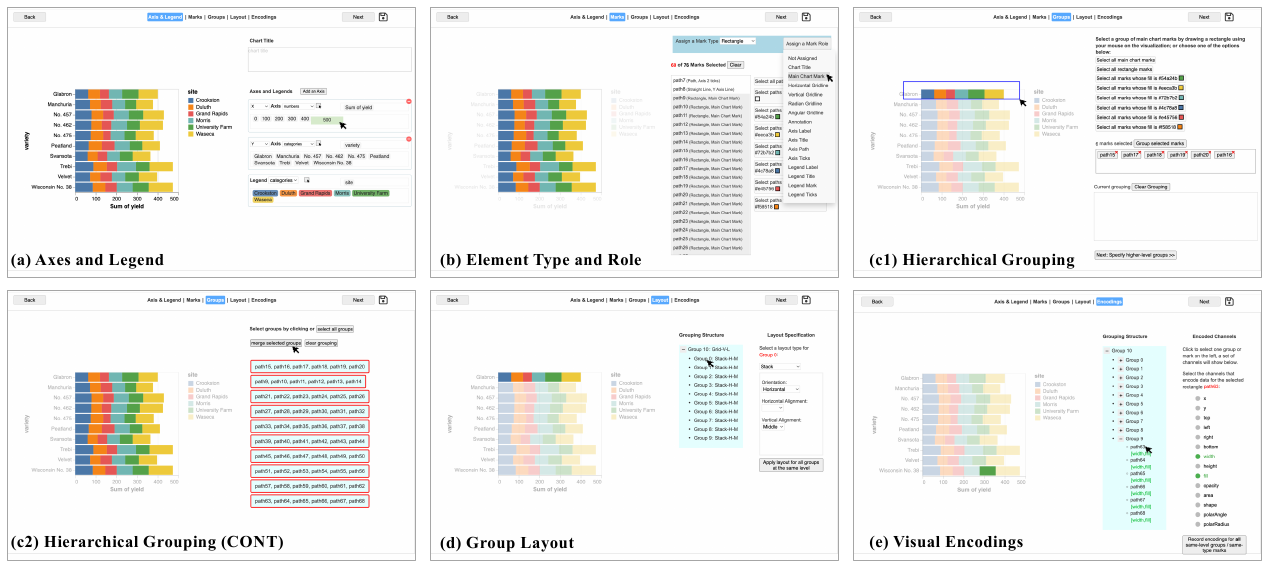}
  \caption{The labeling tool we have developed to produce the semantics labels in \corpus{} divides the labeling process into five stages: \stageAL~(a), \stageM~(b), \stageG~(c1, c2), \stageL~(d), and \stageE~(e) and provides necessary interactions to accelerate the labeling process.
  % \chen{feels a bit too big; not sure if we want to go back to the 2by3 smaller version} \leo{I think the current size is needed for the text to be legible, unless we re-do the screenshots with the font size enlarged.} \hannah{same, and since CHI doesn't have a page limit, we can be liberal with our use of space.}
  }
    \label{fig:labelingTool}
\end{figure*}

\noindent \allEle\ include every geometric shape and text element in an SVG chart~(leaf nodes in the SVG hierarchy). Each element has several general properties, 
% obtained from parsing the SVG file,  
including element ID, element tag name, text content~(if any), and fill color. Each element's bounding box is expressed in absolute coordinates. Figure~\ref{fig:teaser}(c-I) presents three example primitives and their properties in the stacked bar chart. 
% \hannah{this is confusing. I've read this three times and still can't understand what you mean without looking at the figure.}

\noindent  \refEle\ are 
% record the information of 
titles, axes, legends, and gridlines in a chart.
% , specified using IDs of constituent SVG elements.
Titles and gridlines are specified with the IDs of the corresponding SVG elements in the label file.
% \hannah{not sure I understand what this means?}.
Axes and legends contain information about the type (e.g., x, y, angle, radius) and orientation (e.g., horizontal, vertical), and they 
are further broken down into lower-level components such as labels and ticks that are also specified with IDs of their corresponding elements.
% to give detailed representations. 
Figure~\ref{fig:teaser}(c-II) shows the semantic labels for the x-axis, y-axis, and the color legend in the chart. 
% and their detailed element-based information in the label file.

\noindent  \markInfo\ record the shape type and semantic role of each SVG element. In SVG files, the tag name of an element does not always match the geometric shape. \corpus\  resolves this ambiguity through the \markType\ label. For example, in Figure~\ref{fig:teaser}(c-III), ``\texttt{path64}'', the bar representing ``Winsconsin No. 38'' in Waseca, has the tag name ``path'', while its \markType\ is labeled as \texttt{Rectangle}.
% ~(which is different from its tag name \texttt{Path}) and 
 In addition, the same type of elements can play different roles in a chart. For example, elements ``path64'' and ``path82'' are both rectangles in Figure~\ref{fig:teaser}(c-III), but the former is labeled with the \markRole{} \texttt{Main Chart Mark} and the latter is labeled as a \texttt{Legend Mark}. 
 % \hannah{is there a list of all the roles an element can have? link?}
 % the element  ``\texttt{text0}'' represents the zero value in the x-axis and is labeled with \texttt{Axis 1 Label} as its role. 
 
\noindent  \groupInfo\ reflects the multi-level semantic clustering of main chart marks~(elements whose \markRole\ is \texttt{Main Chart Mark}, i.e., the 60 colored bars from the main chart area).
% \hannah{this wording is confusing}
The example stacked bar chart has 10 mark groups~(``\texttt{Group1}'' to ``\texttt{Group10}''), each corresponding to one of the 10 varieties of barley~(i.e., ``Glabron'' to ``Winsconsin No. 38'' along the y-axis), and they further form a higher-level group, ``\texttt{Group11}'', that encapsulates all the 10 lower-level groups. This hierarchy is recorded in the label file as shown in Figure~\ref{fig:teaser}(c-IV).

\noindent  \layoutInfo\ indicates the spatial relationship between visual objects within one group~(e.g., grid, stack, packing, radial) 
% \leo{Give a few examples to illustrate the idea of layout type}
with orientation~(e.g., horizontal, vertical, angular) and alignment parameters~(e.g., bottom-aligned, left-aligned). 
This information is present for all the groups at all levels. For example, in Figure~\ref{fig:teaser}(c-V), ``\texttt{Group1}'' is labeled with a middle-aligned horizontal stack layout, and the same layout applies across ``\texttt{Group2}'' to ``\texttt{Group10}''; the higher-level ``\texttt{Group11}'' is labeled as a left-aligned vertical grid layout.

\noindent  \encInfo\ records which visual elements and channels are used to encode data.
% for main chart marks and mark groups. 
% \hannah{rewrote this, does it make sense?}
Figure~\ref{fig:teaser}(c-VI) shows labels for visual encodings in the stacked bar chart, indicating that the bar ``\texttt{width}'' and ``\texttt{fill}'' encode data values.
% from the underlying data table. 
In some charts, the visual channels of a group can encode data as well. For example, in a small-multiple design of grouped bar charts, the position of each bar chart can encode the approximate geographic location of the corresponding U.S. state. 
\corpus~thus organizes visual encoding labels by each visual object's ID. 

\revision{These semantic components are used as fundamental building blocks in programming libraries such as Mascot.js \cite{liu2024msc} and authoring tools like Charticulator \cite{ren_charticulator_2018}. They are expressive enough to describe not only standard charts but also bespoke designs, as evidenced by these tools' galleries. Our labeling effort, described in the next section, confirms the expressivity of these components.}

% \leo{We may not need this paragraph.} To summarize, \corpus\ provides rich fine-grained semantic labels of each chart's components, including each graphical element's type (e.g., rectangle, pie, polyline), role (e.g., mark, axis path, legend tick, annotation), properties (e.g., ID, fill color, content) and bounding box, the hierarchical grouping of graphical elements, the layout for each element group, and visual encodings on main chart elements. 
% The richness and multi-granularity of the semantic labels in\corpus\ have the potential to unlock a broad set of visualization applications. \fy{``Show, don't tell'' Show potential and applications; don't just tell people there is potential. If you state in later sections, add a pointer here}
% The \supp{}\footnote{see \href{https://osf.io/962xc/?view_only=adbb315fd8794f6dac6b9625d385900f}{\color{blue}{detailed statistics}} on the semantic labels in \corpus.} provides more statistical analysis of the semantic labels.
% \leo{now we are adding a new section on this?} 
% \fy{I'm a little puzzled. What stats analyses for what purposes? My guess is that you provide distributions of your corpus labels? this isn't very clear}
% We provide details on the labeling process in Section~\ref{sec:labelingProcess}. \fy{Q: how did you decide on these labels? }

% \parHeading{Statistics on the Semantic Labels.} \mytodo{maybe in the supp?} 
% \leo{make this sec 2.2, and the former sec 2.1?}

% the average ratio of \#main chart mark in each type or in the whole corpus

% distribution of number of axes?

% distribution of number of marks/mark types?

% distributions of layouts/encodings?

% distribution of nesting levels?

\begin{table*}[!hb]
\caption{A comparison between \corpus\ and nine related chart corpora in their current states. {\footnotesize \exist} indicates the existence of a certain property in the corresponding corpus, {\footnotesize \partialExist} indicates partial existence, {\footnotesize \canExist} means a property is absent in the current state but can be obtained with some effort, and - indicates that a property is unavailable. 
}
\centering
\resizebox{\textwidth}{!}{
\begin{tabular}{lrcccccccccc}
\toprule
 &                & VisAnatomy & MVV~\cite{mvv} & VisEval~\cite{visEval} & YOLaT++~\cite{dou2024hierarchical} & REV~\cite{poco_reverse-engineering_2017} & VisText\cite{benny2023vistext} & Beagle~\cite{battle2018beagle} & MASSIVE~\cite{massive} & VisImages~\cite{deng_visimages_2022} & Chart-LLM~\cite{ko2023natural} \\ \midrule
\multicolumn{2}{l}{Primary Collection Method} & \shortstack{Manual\\Curation} & \shortstack{Manual\\Curation} & \shortstack{Transforming An\\Existing Corpus} & \shortstack{Computer-Aided\\Generation} & \shortstack{Computer-Aided\\Generation} & \shortstack{Computer-Aided\\Generation} & \shortstack{Web\\Crawling} & \shortstack{Web\\Crawling} & \shortstack{Web\\Crawling} & \shortstack{Web\\Crawling} \\ \midrule
\# Charts   &    & 946 & 360 & 1,150 & 15,197 & 5,125 & 12,441 & \textbf{$\sim$41,000} & 2,070 & 12,267 & 1,981 \\ \midrule
\# Types   &    & \textbf{40} & 14 & 7 & 11 & 4 & 3 & 24 & 12 & 34 & 10 \\ \midrule
\# Tools   &    & \textbf{58} & - & 1 & 2 & - & 1 & 5 & - & - & 1 \\ \midrule
\multirow{3}{*}{Format} 
& SVG    & \exist & - & - & \exist & \partialExist & \canExist & \exist & - & - & \canExist \\ \cmidrule{2-12} 
& Bitmap    & \exist & \exist & - & \canExist & \exist & \exist & \exist & \exist & \exist & \canExist \\ 
\cmidrule{2-12} 
& Program   & - & - & \exist & \exist & - & \exist & - & - & - & \exist \\ \midrule
\multirow{12}{*}{Label} 
& Chart Type    & \exist & \exist & \exist & \exist & \exist & \exist & \exist & \exist & \exist & \exist \\ \cmidrule{2-12} 
& Chart BBox    & \exist & \exist & - & - & - & - & - & - & \exist & - \\ \cmidrule{2-12} 
& Element BBox    & \exist & - & - & \exist & \partialExist & \partialExist & - & - & - & - \\ \cmidrule{2-12} 
& Legend Elements   & \exist & - & - & \exist & \partialExist & - & - & - & - & - \\ \cmidrule{2-12} 
& Axis Elements   & \exist & - & - & \exist & \partialExist & \exist & - & - & - & - \\ \cmidrule{2-12} 
& Element Shape    & \exist & - & - & \exist & - & \partialExist & - & - & - & - \\ \cmidrule{2-12} 
& Element Role   & \exist & - & - & \exist & \partialExist & \partialExist & - & - & - & - \\  \cmidrule{2-12} 
& Mark Grouping    & \exist & - & - & - & - & - & - & - & - & - \\ \cmidrule{2-12} 
& Group Layout   & \exist & - & - & - & - & - & - & - & - & - \\ \cmidrule{2-12} 
& Visual Encoding   & \exist & - & - & - & - & - & - & - & - & \canExist \\ \cmidrule{2-12}
& (NL, VIS) pairs   & - & - & \exist & - & - & \exist & - & - & - & \exist \\
\bottomrule
\end{tabular}
}
\label{table:comparisonWithOtherCorpora}
\end{table*}

\subsection{Labeling with A Semi-Automated Tool}\label{sec:labelingTool}

These semantic labels are created using a semi-automated chart labeling tool we have developed. 
% \parHeading{Labeling with The Semi-automated System} \hannah{this could be a subsubsection} 
According to Chen and Liu~\cite{chen2023state},
% which discusses researchers' common practices and potential pitfalls \leo{in what?}~\cite{chen2023state}, 
obtaining high-quality chart labels is expensive and time-consuming, especially for complex labels that require careful examination of charts. Moreover, SVG charts have diverse hierarchical structures and utilize SVG elements in various ways, especially when they come from different tools and sources. 
% \chen{maybe talk more about the challenges here} 
% Given the promoted diversity of \corpus, it is impractical to perform labeling purely through examining the SVG files. 
To address these challenges, we have developed a mixed-initiative multi-stage labeling tool to facilitate the process and mitigate laborious inspection. 

The annotation tool divides the labeling process into five stages: \stageAL, \stageM, \stageG, \stageL, and \stageE. Figure~\ref{fig:labelingTool} shows the system user interface and an example labeling workflow using the stacked bar chart in Figure~\ref{fig:teaser}(c). 
Upon loading the chart, the system traverses the SVG hierarchy to obtain \allEle. Then the system leverages the heuristics-based methods reported in the Mystique system~\cite{mystique} to detect the chart title, axes, and legend, and displays the results in the \stageAL{} UI accordingly~(Figure~\ref{fig:labelingTool}(a)). The annotator can correct the results through drag-and-drop or lasso selection over SVG texts to revise \refEle; for example, in Figure~\ref{fig:labelingTool}(a), the annotator is dragging the text ``500'' from the chart into the x-axis label box. The annotator can also add more axes if more than two axes exist~(e.g., in parallel coordinates).

Once the annotator finishes inspecting the results in this stage, they can click the ``Next'' button to go to the \stageM{} stage, where the full list of graphical elements is displayed~(Figure~\ref{fig:labelingTool}(b)). The annotator can click to select a single mark, batch-select multiple marks through the generalized selection options~\cite{generalizedSelection}~(e.g., selecting the same-type or same-color marks), or shift-click to select consecutive marks in the mark list. 
% to select a set of marks and batch-process their mark types and roles. 
When a set of marks is selected, they will be highlighted in full opacity in the chart, while all other elements will be partially transparent. 
% also shift-click to select consecutive marks in the mark display panel, for example, in Figure~\ref{fig:labelingTool}(b) the annotator has selected all 60 colored rectangles by shift-clicking and is 
The annotator can label the \markInfo\ of the selected marks through the corresponding drop-down menus. In Figure~\ref{fig:labelingTool}(b), the selected paths are assigned \markType{} \texttt{Rectangle} and \markRole{} \texttt{Main Chart Mark}. At this stage, the annotator would also need to specify gridlines and low-level components for axes and legend~(e.g., ticks, paths), so that the \refEle\ label is complete.

In the next stage, \stageG, the annotator specifies \groupInfo\ on all the main chart marks. \revision{Since the use of the \texttt{<g>} tag is inconsistent across charts produced by different tools \cite{mystique,chen2023state}, the original SVG grouping information is ignored by the labeling tool.} To select marks for grouping, the annotator can click to select individual elements, make a lasso selection (Figure~\ref{fig:labelingTool}(c1)), or choose from the generalized selection options. 
% The system extracts all main chart marks~(elements whose role is \texttt{Main Chart Mark}) and listens to the lasso selection or click events from the annotator. Through these two interactions, the annotator can select a set of marks~(Figure~\ref{fig:labelingTool}(c1)), and mark them as a group 
The selected marks can then be grouped by clicking the ``Group selected marks'' button. After that, the system will recommend a list of inferred ``other mark groups'', and the annotator can either agree to finish the lowest-level grouping or disagree to continue grouping manually. Once all lowest-level mark groups are specified, the annotator clicks the ``Specify higher-level groups'' button to go to the second phase of the \stageG{} UI, 
% ~(Figure~\ref{fig:labelingTool}(c2)), 
where they work on higher-level grouping, e.g., in Figure~\ref{fig:labelingTool}(c2) the annotator has selected all 10 mark groups and are merging them into one final group. 

After \groupInfo\ is finished, the annotator goes to the \stageL{} stage. Here the hierarchical groups will be displayed as a collapsible nested list with clickable items. The annotator can click on an individual group and label its layout type and parameters, as shown in Figure~\ref{fig:labelingTool}(d). In the final \stageE{} stage, the system adds mark items to the nested list and lets the annotator specify their encoded visual channels~(Figure~\ref{fig:labelingTool}(e)). The visual channel list will be updated accordingly based on \markType{} of the selected element. In both the \stageL{} and \stageE{} stages, the annotator can apply the label of one group or mark to its peers, i.e., same-level and same-type objects, to accelerate the process of assigning \layoutInfo\ and \encInfo. 

At any time during the labeling process, the annotator can click on the ``save'' button in the upper-right corner of the interface to save a local copy of all semantic labels created so far in the JSON format. When the same chart is loaded next time, the system will automatically retrieve the corresponding label file (if it exists), and load the semantic labels in the UI accordingly to allow the annotator to inspect existing labels and continue unfinished work. The semi-automated labeling system and a document recording the options for \markType, \markRole, types and parameters of \layoutInfo{}, and channels of \encInfo, are included in the supplementary materials\footnote{see the \href{https://osf.io/962xc/?view_only=adbb315fd8794f6dac6b9625d385900f}{\color{blue}{code and details}} for the labeling system.}. 
% \leo{we should include a list of complete mark types, roles, layout, channels in the supp materials. The codebase has these but it's better to make them more explicit. }

\subsection{Iterative Annotation and Quality Control}
\label{sec:labelQuality}

Using the system to produce high-quality semantic labels requires sufficient knowledge and expertise in the visualization field. 
Thus, we adopt an iterative approach to obtain the labels from three experts who are familiar with visualization abstraction papers in the team. 
The first author, a Ph.D. student who has published peer-reviewed papers in visualization-related conferences~(e.g., VIS, EuroVis), performed the first-round labeling. 
The time required to finish labeling one chart ranges between 10 to 20 minutes.
After that, the second author, a Ph.D. student who has published at visualization-related conferences such as VIS and IUI, and the last author, a researcher who has been contributing to the visualization community for more than 15 years, performed the second-round inspection. 
Each chart has been reviewed by at least two experts on the team.
Whenever disagreements over certain semantic labels arose, the three authors discussed the cases to reach a consensus on the correct labeling approach, and all the charts with features related to the issues would be re-labeled. 
These discussions not only improved the consistency and accuracy of the labels but also enhanced the labeling system.

\input{}

\subsection{Comparing \corpus\ with Related Corpora} 
\label{sec:comparison}

In this section, we compare \corpus\ with nine existing chart corpora: Beagle~\cite{battle2018beagle}, YOLoT++~\cite{dou2024hierarchical}, REV~\cite{poco_reverse-engineering_2017}, MASSIVE~\cite{massive}, MVV~\cite{mvv}, VisImages~\cite{deng_visimages_2022}, Chart-LLM~\cite{ko2023natural}, VisText~\cite{benny2023vistext}, and VisEval~\cite{visEval}~(results are shown in Table~\ref{table:comparisonWithOtherCorpora}). These corpora were selected based on the following criteria: the corpus should be publicly available, contain chart images, and have been published at VIS or HCI venues. 
Based on the criteria, we did not include the Visually29k corpus~\cite{Visually29K} as it focuses on infographics and the VIS30K corpus~\cite{vis30k} which contains images of data tables. 

\corpus\ distinguishes from other chart corpora in two major aspects: the rich, multi-granular chart semantic labels, and its diversity regarding chart types, designs, and sources. In terms of the number of charts, \corpus is comparable to datasets that specifically emphasize label quality, such as VisEval~\cite{visEval} and Chart-LLM~\cite{ko2023natural}, which contain around one or two thousand samples. Although corpora such as Beagle \cite{battle2018beagle} collect many more charts through web crawling, they tend to include duplicate designs with unbalanced chart distributions \cite{mystique}. Moreover, they lack fine-grained component-level labels. On the other hand, corpora created through computer-aided generation (\eg YOLaT++\cite{dou2024hierarchical}) do provide fine-grained component labels, but are highly restricted in terms of chart and tool diversity. \corpus balances between promoting diversity for real-world charts and maintaining quality control for fine-grained labels. The labels for over 380K visual elements in \corpus adequately support applications requiring extensive component-level annotations.

\section{Use Cases}\label{sec:useCases}
To showcase the utility of \corpus{}, we present four use cases, each focusing on a specific downstream visualization application:
\begin{itemize}[leftmargin=*, itemsep=-1pt]
    \item \textit{\revision{Inferring semantic roles of SVG elements}} with Large Language Models~(LLMs): \revision{we evaluate two LLMs to examine their capabilities of classifying SVG elements into component types}. 
    \item \textit{Decomposing} rectangle-based \textit{SVG charts for layout reuse}: we use \corpus~to obtain a validation set to evaluate the performance of an existing system Mystique~\cite{mystique}.
    \item \textit{Classifying chart types} using Graph Neural Networks~(GNNs)~\cite{xu2018how} \revision{and vision models: we report and compare the performance of vision models and graph models using \corpus~as a benchmark corpus, and discuss the trade-offs involved}.
    \item \textit{Supporting accessible navigation} of chart content for visually impaired people: we demonstrate how \corpus~can enable the replication of the rich screen reader experiences through keyboard navigation as described in the study by Zong et al.~\cite{screenReader}.
\end{itemize}

% \fy{I'm thinking: I'd love to read use cases!}

The four use cases span different requirements on the semantic labels, from low-level mark details~(e.g., SVG element role) to high-level chart information~(e.g., chart type). Through these applications, we demonstrate how \corpus~can enable tighter integration of AI in visualization. 
% its role in introducing new paradigms for AI in visualizations. 
In addition, we illustrate how the semantic labels in \corpus{} ease the constraints on input formats such as the necessity for charts to be created with specific tools, supporting a wider range of input charts. The \supp{}\footnote{see our \href{https://osf.io/962xc/?view_only=adbb315fd8794f6dac6b9625d385900f}{\color{blue}{implementations, results, and demo videos}} for the use cases.} include our implementations, detailed results, and demo videos for the use cases. 
% \fy{Wow!}

% 

\subsection{\revision{Semantic Role Inference with LLMs}}
\label{sec:llm4path}

\revision{
Inferring chart semantics is a classic task in automated visualization understanding. For instance, ReVision~\cite{savva_revision_2011} utilizes vision models to detect the mark types and the underlying data used in a bitmap visualization. More recently, with the rapid advancement of LLMs, researchers have found that (1) text-based chart specifications improve the performance of language models (LMs) on chart-reading tasks compared to vision-based approaches~\cite{victor2024representing}, and (2) the existence of semantic information for SVG elements (so called ``Primal Visual Description (PVD)'')  boost the performance of LMs for vector graphics reasoning~\cite{wang2024text}. 
In this use case, we would like to examine the  capabilities of current LLMs in semantic inference; more specifically, we focus on the task of classifying SVG elements into the following categories: main chart elements (marks), axes (titles, labels), and legend (titles, labels, marks).  This classification is an important prerequisite for many downstream tasks~\cite{mystique,poco_reverse-engineering_2017}, and we evaluate it using one open-source LLM~(DeepSeek-V3-0324~\cite{liu2024deepseek}) and one proprietary LLM~(GPT-4o~\cite{hurst2024gpt}).
% , which serves as an important step of many downstream tasks~\cite{poco_reverse-engineering_2017,mystique}.
% In this use case, we generate a large-scale dataset of SVG \textit{path} elements with ground-truth mark types based on \corpus, and evaluate the shape recognition capability of three open-source LLMs: Llama-3.1-8B-Instruct~\cite{dubey2024llama},Mistral-7B-Instruct-v0.3~\cite{llm_Mistral}, and DeepSeek-V2.5~\cite{deepseekv2}.
}

% \parHeading{Dataset of SVG \textit{path} elements.} We first select all the SVG elements with the tag ``path'' and their corresponding shape labels (i.e., \markType), forming an initial dataset of 175,065 path elements.
% Since the annotations in \corpus include the real shape  of all \texttt{Main Chart Marks}, we obtain those whose SVG tag is \textit{path} and record their real shapes to form an initial \textit{path} dataset, which consists of $175,065$ \textit{path} elements. 
% However, this dataset is highly unbalanced because \textit{rect} marks account for more than $100$k of them. We therefore randomly sampled up to $1,000$ elements per shape type (or all available elements if fewer than $1,000$ existed) to form the final dataset which includes $9,914$ \textit{path} elements. This dataset encompasses ten shape types: Straight Line, Polyline, Rectangle, Polygon, geoPolygon, Circle, Pie, Arc, Area, and Text. Among these, Circle is the only shape type with fewer than $1,000$ elements ($914$).

\parHeading{Prompt.} \revision{The prompt contains the role specification (``\textit{expert in analyzing SVG-based data visualizations}''), the task (``\textit{identify and categorize different visual elements in the provided SVG chart into main chart marks, legend components (title, mark, label), and axis components (title, label)}''), explanations on the semantic categories, and the required JSON-format output. 
% The supplementary materials contain the detailed prompt. 
For both models, we utilized their cloud APIs for inference due to their significant large model sizes.
}

% \parHeading{Experimental Environments.} We conducted inference using different hardware configurations based on model size. For Llama-3.1-8B-Instruct and Mistral-7B-Instruct-v0.3, inference was performed locally on a single NVIDIA RTX A5000 GPU. For DeepSeek-V2.5, due to its significantly larger model size, we utilized its cloud API for inference.

\begin{figure}[!ht]
  \centering
  \includegraphics[width=0.49\textwidth]{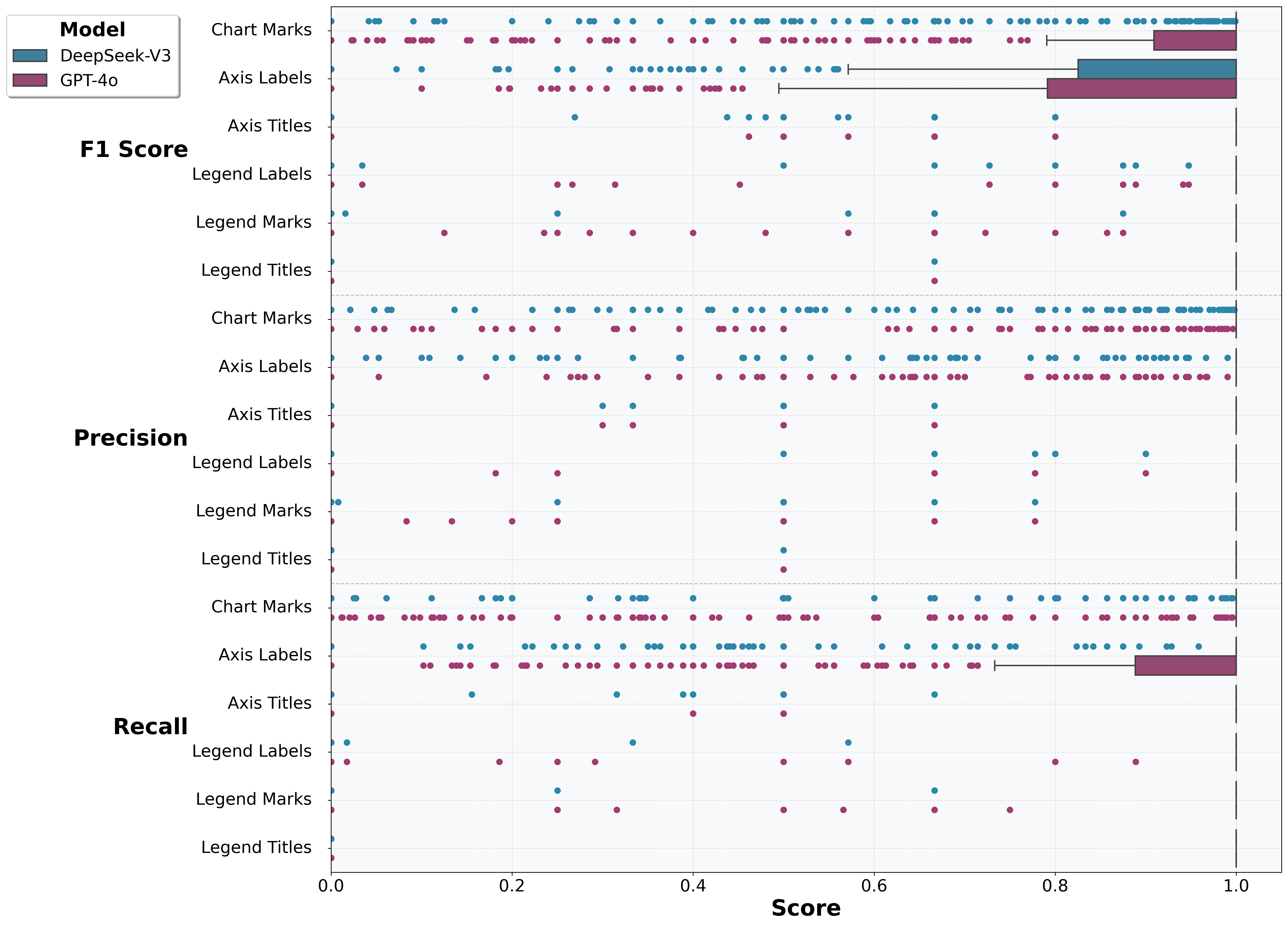}
  \caption{Performance of two LLMs in inferring the semantic roles of SVG elements for 772 charts. The whiskers show 1.5~$\times$~IQR thresholds, and the dots are considered ``outlier'' charts.}
    \label{fig:llm4semantics}
    \vspace{-3mm}
\end{figure}

\parHeading{Results.} \revision{
DeepSeek-V3 successfully processed 772 charts and GPT-4o processed 835, according to their respective context length constraints (i.e., input token limits of 64K for DeepSeek-V3 and 128K for GPT-4o).
To make it a fair comparison, we report results from the 772 charts processed by both models.
% and some raw SVG file being extremely long, at the end . 
We compare their inferences with the ground truth annotations from \corpus~(\texttt{Main Chart Marks} and \refEle), and record F1, precision, and recall scores for each semantic category. }

\revision{
The two LLMs achieve comparable performances in terms of the overall F1 scores: DeepSeek-V3 (mean $0.841$, median $0.870$), GPT-4o (mean $0.839$, median $0.929$).
% ($0.84$ and $0.839$ respectively), with DeepSeek-V3 being slightly better. 
% The median scores for F1, precision, and recall are 1 in all categories for both models. 
Both models performed well in recognizing \textit{axis titles}, \textit{legend marks}, \textit{legend labels}, and \textit{legend titles}, with the maximum, median, and 1.5~$\times$~IQR all equal to 1 (\Cref{fig:llm4semantics}). 
% some outliers shown as dots in the box-and-whisker plots . 
The performance of both models in inferring \textit{main chart marks} and \textit{axis labels} is comparatively lower than for the other categories, with DeepSeek-V3 demonstrating superior results to GPT-4o for these two categories. 
% Regarding the specific semantic categories, the two models have close performances for axis components, legend labels and legend marks. When it comes to legend title, GPT-4o has a clear advantage where its mean F1 score is $13.7\%$ higher than that of DeepSeek-V3~($0.945$ versus $0.831$); while DeepSeek-V3 performs better in recognizing main chart marks with a $8.4\%$ advantage ($0.916$ versus $0.845$). 
Based on this result, \textit{the current state-of-the-art LLMs show promising performance in recognizing the semantic roles of SVG elements}, indicating a great potential to develop LLM-assisted systems for SVGs. In addition, \textit{when working with these LLM inferences, careful human verifications and corrections need to be supported, especially for main chart marks and axis labels.}
% It should also be noted that, as shown in \Cref{fig:llm4semantics}, the standard deviation values are relatively high~(ranging from $0.2$ to $0.4$); as a result, \textit{when working with these LLM inferences, careful human verifications and corrections need to be supported}.
}

\revision{
We also examined model performance across different chart types, as these vary in design elements and complexity.  The key finding here is that \textit{GPT-4o tends to be more accurate on more complex charts compared to DeepSeek-V3, and vice versa}. For instance, DeepSeek-V3 achieves better mean F1 scores in traditional statistical visualizations such as simple bar charts (0.887 vs 0.543, 63\% performance gap), area charts (0.876 vs 0.641, 37\% gap), and stacked bar charts (0.902 vs 0.607, 49\% gap), while GPT-4o exhibits superiority with lower margin in complex visualizations including circle packing diagrams (0.959 vs 0.885, 8\% gap), bullet charts (0.850 vs 0.697, 22\% gap), and bespoke visualizations (0.915 vs 0.846, 8\% gap). These performance gaps may be attributed to potential differences between visualization training data used for these LLMs and suggest the need for fine-tuning for better results across chart types.
}

\revision{
Finally, we analyzed model performance across charting tools, focusing on 18 tools that each contribute at least 10 charts, resulting in a total of 401 charts. (Note that tool information is unavailable for certain charts in \corpus.) DeepSeek-V3 achieves a mean F1 above 0.9 for 16 of these tools, while GPT-4o does so for only 7 tools. AnyChart \cite{AnyChart}, FlexChart \cite{FlexChart}, AngularChart \cite{AngularChart}, Matplotlib \cite{Matplotlib}, Semiotic \cite{Semiotic}, and Mascot.js \cite{liu2024msc} appear in the top performers for both models. The performance on charts created with the following tools shows a substantial discrepancy between the two models: R (0.692 for GPT-4o, 0.959 for DeepSeek-V3) and NIVO \cite{nivo} (0.765 for GPT-4o, 0.960 for DeepSeek-V3). D3.js \cite{bostock_d3_2011} presents a challenging case for both models (0.769 for GPT-4o, 0.863 for DeepSeek-V3), probably due to D3's high versatility and expressivity.   
}

\subsection{Chart Layout Deconstruction}

Unlike previous corpora that lack semantic information on low-level components, \corpus\  specifies the mark grouping, layout, and data-channel encoding information, which play a vital role in the task of chart decomposition and reuse. Most approaches that focus on this task, such as D3 Deconstructor~\cite{harper_deconstructing_2014} and ChartReuse~\cite{cui_mixed-initiative_2022}, require the input charts to be created using a specific tool. A more recent work, Mystique~\cite{mystique}, designed a tool-agnostic decomposition and reuse pipeline for general SVG charts composed of rectangle-shape marks based on a diverse 150-chart corpus. What lies at the core of Mystique is a bottom-up hierarchical clustering algorithm that determines nested mark groups and their internal spacial relationships presented in a chart. In this section, we use \corpus\ to form a validation set to evaluate the generalizability of the hierarchical clustering algorithm from Mystique on unseen charts. 
% and discuss...\leo{to fill}

To prepare the validation set, we first filter the charts based on the \markType\ labels, and only keep those having the \texttt{Rectangle} type for all \texttt{main chart marks}, resulting in 306 charts.
% \chen{148/462 = 32.03\% which is pretty close to the statistics in Table 1 from Mystique}. 
% Since Mystique is trained and tested on 150 SVG charts, w
Given that Mystique does not consider radial and spiral layouts, we exclude spiral plots and bar charts in the radial layout.
We further remove charts that were already included in Mystique's corpus.
The final validation set consists of 248 charts encompassing 15 chart types. We then run Mystique's hierarchical
clustering algorithm using the \texttt{main chart marks} of each chart in this validation set as input, 
% to be specific, the main chart marks of a given chart are made the input for the algorithm
and the results are compared with \groupInfo\ and \layoutInfo\ labels in \corpus. For the error cases, we perform another round of manual inspection to avoid false negatives as some charts have multiple reasonable grouping structures~(e.g., a diverging bar chart)~\cite{mystique}. 

\begin{figure}[ht]
  \centering
  \includegraphics[width=.495\textwidth]{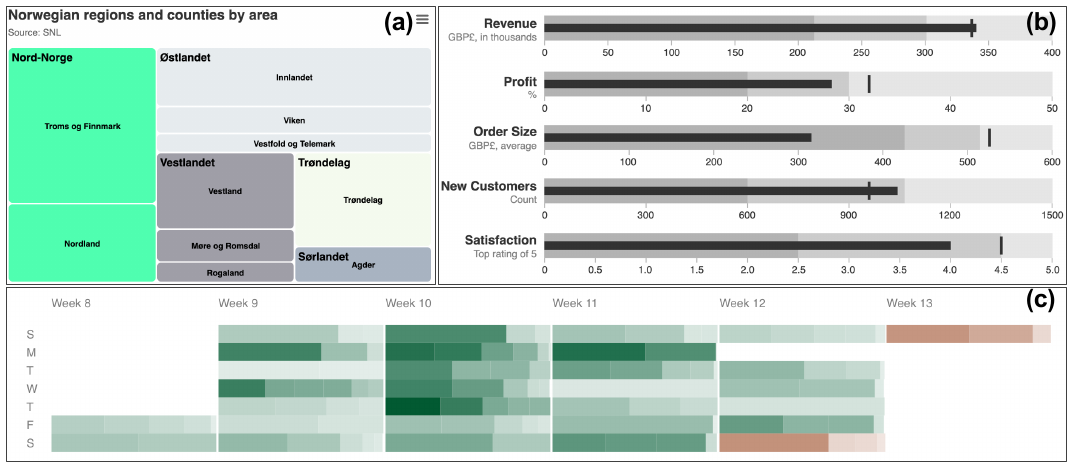}
  \caption{Three new kinds of failure cases have been observed when evaluating the hierarchical clustering algorithm in Mystique~\cite{mystique} on a validation chart set generated from~\corpus: (a) a treemap visualization, (b) a bullet chart, and (c) a bespoke Gantt-calendar chart.}
  % \hannah{is it possible to include the annotations for these charts as well?}\chen{i am afraid the space won'r allow}
    \label{fig:mystique_errors}
\end{figure}

\parHeading{Results.} $217$ charts (out of 248) within the validation set are decomposed into their correct grouping structures and corresponding spatial relationships, making an approximate $87.5\%$ accuracy~($8\%$ lower compared to the test accuracy reported in Mystique~\cite{mystique}). Examining the 31 error cases, we report three kinds of failure cases that were not discovered by the authors of Mystique~\cite{mystique}:

% \hannah{nice!!}:
\noindent
1. A slice-and-dice treemap~(Figure~\ref{fig:mystique_errors}(a)), whose overall packing layout was wrongly recognized as a vertical stack layout by Mystique's decomposition algorithm, leading to incorrect mark groups;\\
2. Glyph-based charts where the marks within a glyph do not always overlap, e.g., in~Figure~\ref{fig:mystique_errors}(b), the three lighter gray bars are stacked within each row without overlapping. The decomposition algorithm uses overlapping relationships to detect glyphs and fails in such cases;\\
3. A bespoke Gantt-calendar chart~(Figure~\ref{fig:mystique_errors}(c)) where the bar groups are positioned based on data and the bars within each group follow a horizontal stack layout. The algorithm fails to decompose this chart correctly, giving a lowest-level packing layout instead.

% \chen{more research on general charts consisting of more types of marks}

In summary, the generalizability of Mystique's hierarchical clustering algorithm is satisfactory, which corroborates the authors' claim on their corpus' diversity. 
% highlighting the benefit of corpora with promoted diversity. 
Nevertheless, the new error cases and Mystique's lack of support for additional  mark types and layouts call for more intelligent decomposition and mixed-initiative reuse approaches.
% that allow human inputs to fix mistakes in model results. 
% The supplementary materials include details on the validation set and the performance statistics.

% \fy{this section isn't formatted as good as the previous one; add some section headers}

\subsection{Chart Type Classification}\label{subsec:GNNcase}

Chart type classification is a typical visualization task, serving as the first step in many end-to-end systems such as Revision~\cite{savva_revision_2011}, ChartSense~\cite{jung_chartsense_2017}, and REV~\cite{poco_reverse-engineering_2017}. Various models, including Support Vector Machines (SVMs)~\cite{svm} and Convolutional Neural Networks (CNNs)~\cite{cnn}, have been explored. However, existing work mostly assumes bitmap images as inputs, and \revision{focuses on coarse chart taxonomies with approximately a dozen categories. Considering that SVG is a highly structured data format, Graph Neural Networks~\cite{gnn, xu2018how} could be a good fit~\cite{chen2023state}, which have shown better results than image feature-only models in tasks such as chart retrieval~\cite{structure2022Li} on a corpus of Plotly \cite{Plotly} charts. In this case study, we explore two questions: 1) how well GNNs classify SVG charts annotated with various semantic labels, and 2) how vision models and graph models perform when the number of chart types increases to 40.}  

% little is known about the implications of the SVG format on chart classification tasks.
% \hannah{also restate how your approach differs from the beagle work.} \leo{Beagle does that, but not with GNN.} 
% In text-based XML representations, SVG charts cannot be directly processed by vision models. Considering that SVG is a highly structured data format, Graph Neural Networks~\cite{gnn, xu2018how} could be a better fit~\cite{chen2023state}. In a recent work focusing on chart retrieval~\cite{structure2022Li}, GNN is employed to obtain semantic embeddings to calculate pair-wise similarities between charts and is shown to be a more effective approach compared to image-feature-only models. To our knowledge, no previous research has conducted SVG chart-type classification with GNNs. In this case study, we introduce how we prepare the graph data based on the SVG charts and present our preliminary results on employing GNNs to recognize chart types on vector graphics charts. The main goal of this case study is to understand the influence of different types of semantic labels from \corpus\ on the classification results.
% \fy{I think someone told me a study about if GPT can recognize SVG charts at some point but i don't remember who was that nor the title}

\parHeading{Node Feature Extraction and Inclusion Criteria.} If we focus on the raw SVG representations, 
we can designate basic shape elements such as \texttt{<line>} and \texttt{<circle>} as nodes in the graph. These elements correspond to \allEle\ in \corpus, and contain \texttt{main chart marks}, \refEle, and background noise (e.g., background rectangles, offscreen tooltips). 
% Given the vastly different property sets of different SVG elements,
To extract node-level features, we consider the following common features shared by different SVG element types for simplicity:
% Our decided node feature contains the following three parts: 
(1) node-type, which is a one-hot encoding over the SVG element types, (2) node-position, which is a four-dimensional vector indicating the node's bounding box in \texttt{top}, \texttt{right}, \texttt{bottom}, and \texttt{left} coordinates, and (3) node-style, a three-dimensional binary vector indicating the existence of the \texttt{fill}, \texttt{stroke}, and \texttt{stroke-width} style properties in the SVG file. 
We can scale the node-position feature using the width and height of the source SVG chart so that all values are within the range [0, 1].
% to be consistent with other features. 
In addition to shape elements, we also designate 
% When forming the final graph, 
SVG container elements like \texttt{<g>} and \texttt{<SVG>} 
% will be added 
as graph nodes, which only have the node-type feature with other feature dimensions filled with 0. 

With the semantic labels in \corpus, we can also restrict graph nodes to elements with \texttt{Main Chart Mark} as their \markRole, so we only focus on the chart content and remove potential noise. In addition, the node-type feature from the raw SVG file can be inaccurate, as it is normal for SVGs to represent different shapes~(rectangles, circles, \etc) with \texttt{<path>} elements~\cite{chen2023state}. Thus, a more accurate version of the node-type feature can be the one-hot encoding over the ground-truth mark types from \markType.

\parHeading{Edges Definitions.} 
Once the graph nodes are processed, we consider two potential ways to define edges. First, we can add edges based on the hierarchical organization between SVG elements and their parent containers in the raw SVG file. Alternatively, we can add the groups from \groupInfo~(instead of the SVG container elements) as nodes to the graph, and construct edges based on the relation of subordination in \groupInfo.

% Since the full list of SVG nodes contains reference objects and noisy elements~\cite{chen2023state}, another way of forming the graph is to 
% we only include the main chart marks~(obtained through \markRole) and the SVG container elements as the graph nodes, the edge definition can be still based on the hierarchy information in the raw SVG file. In this case, the graph is more compact because reference elements and noise are removed. Thirdly, we only include the main chart marks~(obtained through \markRole) and the groups from the \groupInfo\ labels as the graph nodes, and create edges based on the \groupInfo\ information. 
% and draw edges by tracing their ancestors back to the root SVG node. In this way, we achieve a more compact graph representation and don't lose much semantics information as (1) the main chart marks carry the design idea the most and (2) noisy elements are removed.

% We can further embed the \groupInfo\ into this main-chart-mark graph by replacing the parent-child edges from the SVG with the group-child edges, forming a more meaningful hierarchy.

\parHeading{Graph Construction.}
Combining the above variations of node definition, feature extraction, and edge definition, we present the following four graph representations with increasing amounts of semantic labels:
% with increased pre-knowledge requirements on the SVG charts:
\begin{itemize}[leftmargin=*, itemsep=-1pt]
    \item \textsf{\small SVG-Only}~(\gone): SVG shape and container elements as nodes, and parent-child relationships from the SVG hierarchy as edges;
    \item \textsf{\small SVG-MainChart}~(\gtwo): main chart marks based on \markRole\ and SVG container elements as nodes, and parent-child relationships from the SVG hierarchy as edges;
    \item \textsf{\small SVG-MainChart-MarkType}~(\gthree): main chart marks based on \markRole\ and SVG container elements as nodes, with ground-truth one-hot encoding of \markType\ as the node-type feature, and parent-child relationships from the SVG hierarchy as edges;
    \item \textsf{\small SVG-MainChart-MarkType-Grouping}~(\gfour): main chart marks based on \markRole\ and groups from \groupInfo\ as nodes, with ground-truth one-hot encoding of \markType\ as the node-type feature, and group-child relationships from \groupInfo\ as edges.
\end{itemize}

\parHeading{\revision{Tasks.}} \revision{We consider two classification tasks: a 6-category problem where we follow the taxonomy from VisImage~\cite{deng_visimages_2022} to generate 6 high-level categories~(Area, Bar, Circle, Line, Point, Grid\&Matrix) covering 28 chart types, and the full 40-type problem. Within each type, we split charts into the train and test sets randomly using a $6:4$ ratio.
}

\parHeading{\revision{Model Training.}} \revision{We use a 3-layer Graph Transformer~\cite{yun2019graph} network with $1.37M$ parameters to perform graph-based inference with \gone\ to \gfour. We also finetune on two pretrained vision models: MobileNet-v3-small ($2.55M$ parameters)~\cite{howard2019searching} and ViT-small ($22.1M$ parameters)~\cite{caron2021emerging} with the bitmap images in \corpus. For each model, we perform 5 independent runs to minimize noises; during training, we use the Adam optimizer~\cite{adam} with a learning rate of $0.001$ for graph models and $0.0001$ for vision models for $200$ epochs. The machine was NVIDIA GeForce RTX 2080Ti.}
% The \supp{} include the code for and model training.

% \begin{figure}[t]
% \centering
% \begin{tabular}{cc}
%     \includegraphics[width=0.23\textwidth]{figs/training_results_test_acc_c6.pdf} &
%     \includegraphics[width=0.23\textwidth]{figs/training_results_test_acc_c40.pdf} \\
%     \small (a) 6-class classification task. & \small (b) 40-class classification task. \\
% \end{tabular}
% \caption{Learning curves from the six models: top-1 test accuracy.}
% \label{fig:GNN_learning_curves}
% \end{figure}

\begin{figure}[t]
\centering
\begin{tabular}{c}
    \includegraphics[width=0.9\columnwidth]{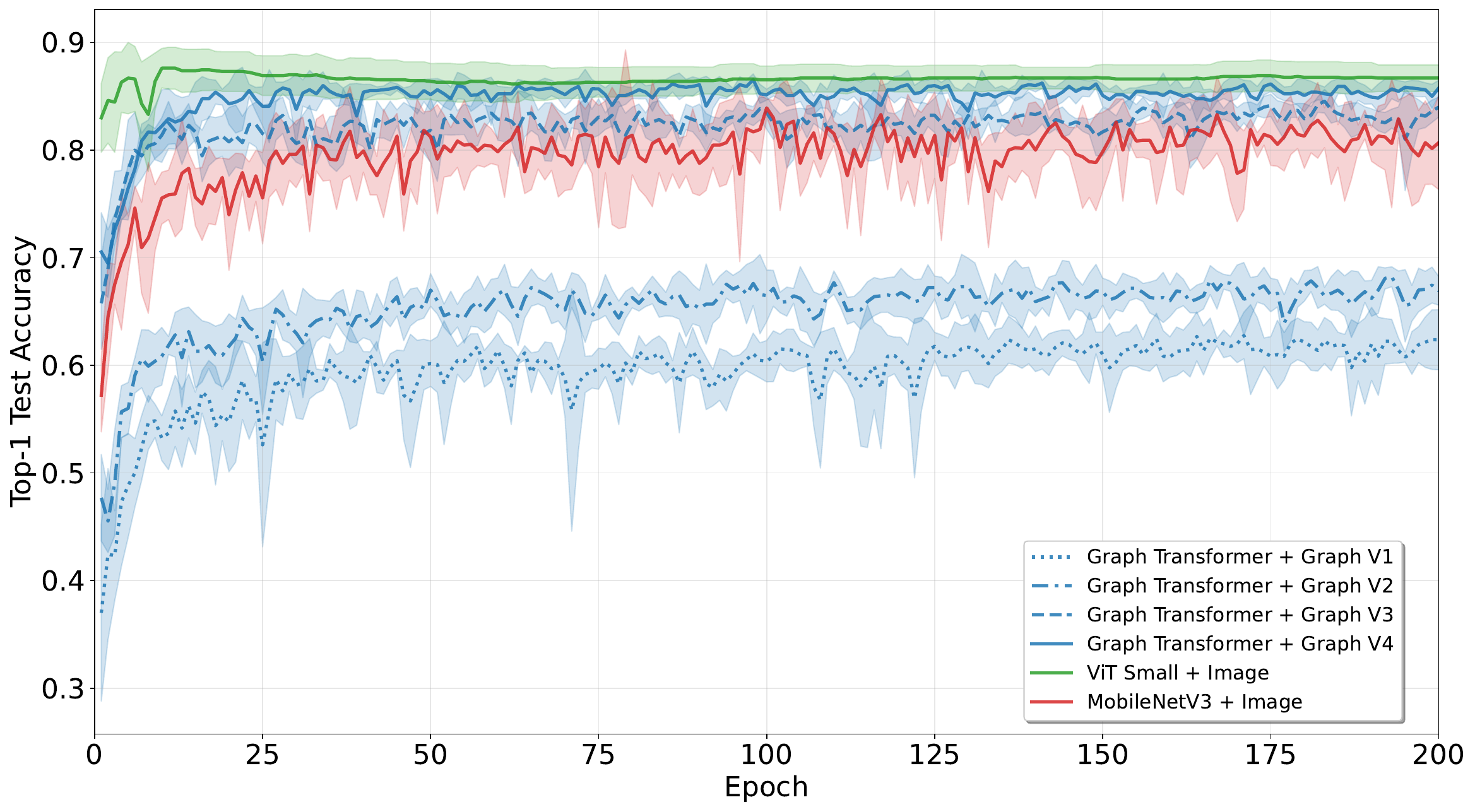} \\
    \small (a) 6-class classification task. \\
    \includegraphics[width=0.9\columnwidth]{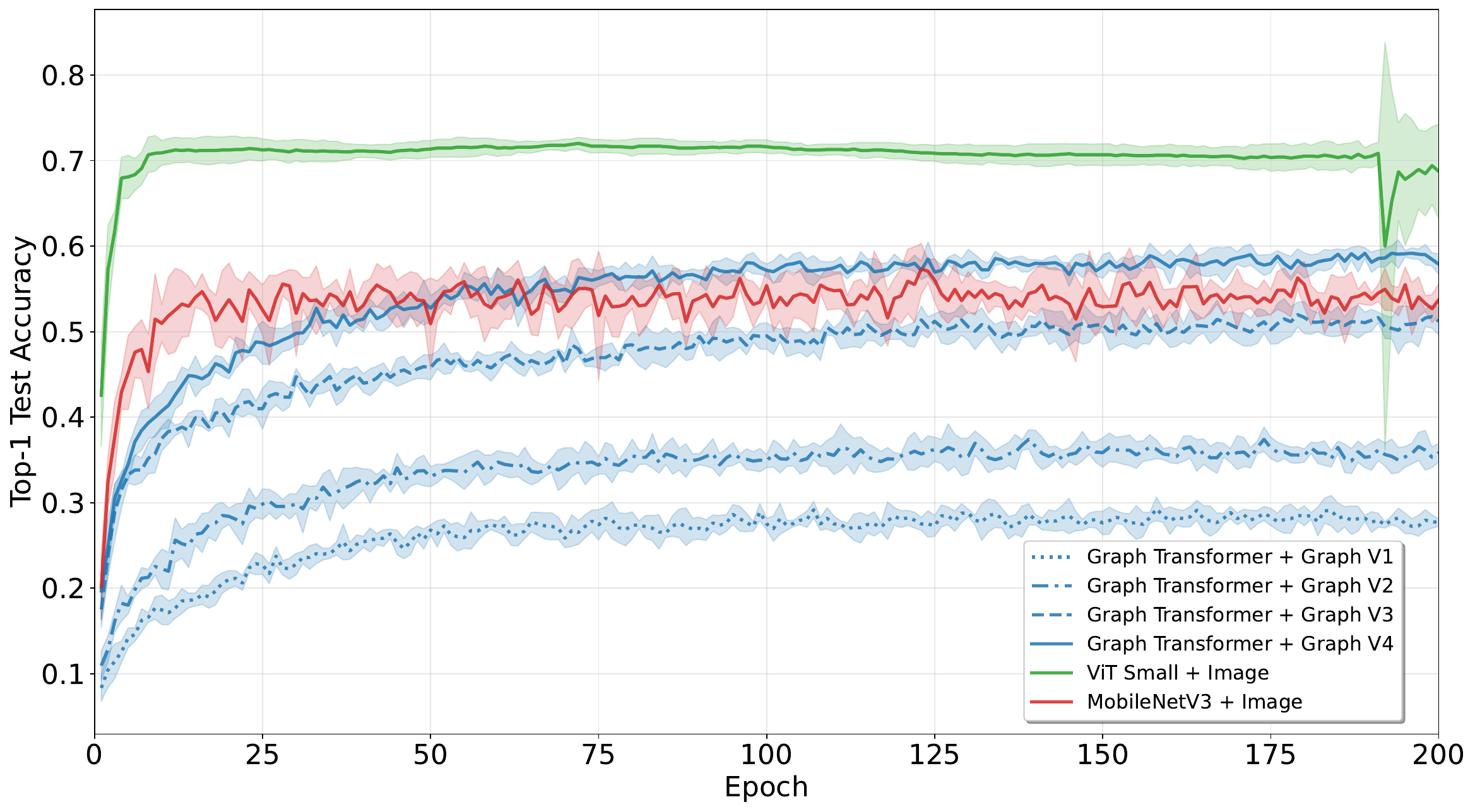} \\
    \small (b) 40-class classification task. \\
\end{tabular}
\caption{Learning curves from the six models: top-1 test accuracy.}
\label{fig:GNN_learning_curves}
\vspace{-3mm}
\end{figure}

\parHeading{\revision{Results.}} \revision{Figure~\ref{fig:GNN_learning_curves} presents the learning curves from the six models regarding the top-1 test accuracy. Generally, in both tasks we can observe that with more semantic information revealed to Graph Transformer, its performance is significantly enhanced (especially from \gthree\ to \gfour), demonstrating the usefulness of \markRole, \markType, and \groupInfo{} from \corpus.}

\revision{Compared to MobileNet-v3 which has $86\%$ more parameters, Graph Transformer obtains the same-level performance with \gthree\ and better test accuracy with \gfour. Although semantic labels are required prior to training, the Graph Transformer can be trained much faster: its average second-per-epoch is $0.53$ while that for MobileNet-v3 is $30.8$. The best performance is given by ViT-small, which has approximately $16.7\%$ gain in test accuracy compared to Graph Transformer with \gfour\ for the 40-class task. However, ViT-small is a much larger model with 15 times more parameters and an average second-per-epoch of $33.37$. Based on the result, test-time label inference and semantic-rich graph construction for enhanced graph-based modeling are promising directions for future work.
}

% \parHeading{Summary.} Our results demonstrate that more semantic labels lead to increased performance of GNNs, indicating the possibility of developing effective mixed-initiative applications powered by GNNs. 
% Because our contribution here is an exploratory study of how the semantic labels can help form graph data and which are useful~(instead of providing well-trained models), there still exist many research questions to explore, such as how to embed more semantics into the nodes and what is the best training strategy. We include more discussions in Section~\ref{sec:discussion}.

% \fy{this use case is convincing}

\subsection{Content Navigation for Accessibility }
\corpus not only serves as a benchmark dataset for evaluating algorithms and models, but also as a valuable resource for researchers to develop visualization applications that use SVG charts as input. We demonstrate its utility by replicating a project focused on chart accessibility, which involves designing chart reading experiences for people with visual impairments. Current visualization accessibility practices link textual descriptions of charts~(usually provided by the authors through alt texts) to the underlying data tables so that assistive screen readers can communicate some high-level chart semantics to the user~\cite{accessibilityExample}. However, this paradigm does not offer visually impaired individuals the same level of chart exploration experiences that sighted people can access through interactive visualizations~\cite{screenReader}. In response, Zong ~\etal~\cite{screenReader} propose a chart accessibility tree as the underlying structure for traversal of a chart's scene graph: the user navigates along multi-level branches of the accessibility tree through a keyboard.
% , \eg pressing the \textbf{Down} arrow key to go to the lower-level element and pressing the \textbf{Left} and \textbf{Right} arrow keys to visit sibling elements in the same level. 

\begin{figure}[!ht]
  \centering
  \includegraphics[width=0.49\textwidth]{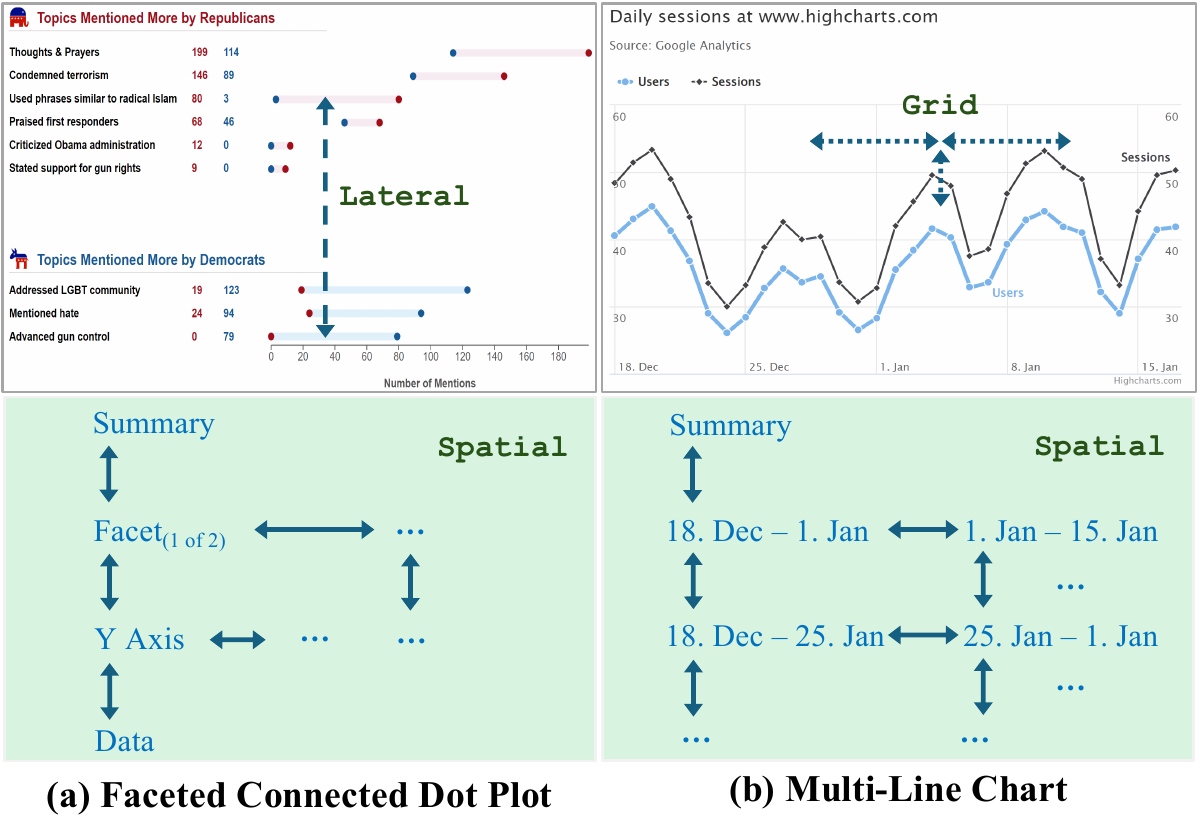}
  \caption{Navigation patterns re-created in (a) a faceted connected dot plot and (b) a multi-line chart using labels from \corpus.}
    \label{fig:screen_5charts}
\end{figure}

Considering that the demos provided by Zong \etal~\cite{screenReader} are implemented with Vega-Lite charts, here we demonstrate that the semantic labels in \corpus\ can support the construction of the accessibility trees for charts created using other tools. 
To re-create the accessible chart reading experiences, we focus on the five examples in the gallery of Zong ~\etal's work~(Figure 2 in \cite{screenReader}), and choose five charts from \corpus~ that have very close visualization designs: a faceted connected dot plot~\cite{ConnectedDot4Access} from Mascot.js, a multi-line chart~\cite{multiLine4Access} from HighCharts, a geographical heatmap~\cite{geoheatmap4Access} from D3.js, 
a stacked bar chart~\cite{stackedBar4Access} from Apexcharts.js, 
and a bar chart with annotations~\cite{annotatedBar4Access} from \textit{poetrybetweenpain.deb.is}. 
% Similar to Zong \etal~\cite{screenReader}, we assume the underlying datasets are available during the construction of accessibility trees.
% \leo{add a sentence such as "Similar to Zong et al., we require the underlying dataset in the construction of accessibility trees"?} 
In our implementation, we focus on the following navigation patterns: 
% presented in~\cite{screenReader}, including 
 \texttt{structural navigation} with the \textbf{Up}, \textbf{Down}, \textbf{Left}, \textbf{Right} arrow keys,  \texttt{spatial navigation} with the \textbf{WASD} keys across grids, and  \texttt{lateral navigation} across facets with the \textbf{Shift+Left} and \textbf{Shift+Right} key combinations.
 % to demonstrate the generalizability of our semantic labels from \corpus, 
 % we resemble the five examples in the gallery of ChartReader~(Figure 2 in \cite{screenReader}) with five charts 
We next briefly introduce how we implemented the accessibility trees~(illustrated in Figure~\ref{fig:screen_5charts}) for the faceted connected dot plot and the multi-line chart example. We include the construction processes for the other three charts in the \supp.

\parHeading{Faceted Connected Dot Plot}~(Figure~\ref{fig:screen_5charts}(a)):  Starting with the whole SVG chart as the root, we introduce two branches corresponding to the two facets using \groupInfo. Each branch has several connected dot pairs as children, which are linked to their corresponding y-axis labels in \refEle\ based on their $y$ coordinates obtained from \allEle. Once the \textbf{Shift + Left/Right} combination is detected, the focus will be shifted to the other facet branch's dot pair of the same index to support the \texttt{lateral navigation}. The leaf nodes in the \texttt{structural navigation} tree are the individual dots. 

\parHeading{Multi-Line Chart}~(Figure~\ref{fig:screen_5charts}(b)): A binary accessibility tree similar to that in~\cite{screenReader} is formed. The range of the x-axis labels from \refEle\ is split into two, each is attached to the SVG root node as a child and linked with marks whose $x$ coordinates are in the range. This binary range partition continues until only one mark is inside; the whole binary accessibility tree is then used for the \texttt{structural navigation}. We further support \texttt{spatial navigation} with the \textbf{WASD} keys across grids: $x$ and $y$ coordinates of gridlines from \refEle\ are obtained to cut the coordinate system into 2-dimensional grids, which are regarded as children of the SVG root. The \textbf{S} key triggers the \texttt{spatial navigation} mode starting with the upper-left grid and the \textbf{Up} arrow key brings the user back to the root.

Overall, we find it straightforward to construct the chart accessibility tree using the semantic labels from \corpus: once the tree structure is decided, the mappings between the tree nodes and the graphical objects in the chart can be retrieved with the help of \groupInfo, \markInfo, position and color properties in \allEle, and axis and legend information from \refEle. 
The \supp{} include demo videos and a web application for re-created navigation patterns on the five charts.
However, we have also observed a few places where semantics beyond \corpus\ are required. For example, 
for the geographical heatmap example, the underlying CSV data from the source website is needed to obtain names of provinces and cities represented by marks; 
% \hannah{what does this mean?}; 
for the annotated bar chart example, the affixation of annotation texts onto the main chart bars needs to be determined in advance. 
We include more discussion on these issues in Section~\ref{sec:discussion}.
% Note that the \texttt{targeted navigation} mentioned in~\cite{screenReader} can also be realized straightforwardly by listing ...
% \hannah{I really love the flow of this section.} \chen{thanks!}

\section{Discussion and Future Work}\label{sec:discussion}
\parHeading{Dealing with Real-World SVG Charts.} SVG charts found in the wild exhibit considerable noise and heterogeneity, even among charts of the same type. The semantic labels in \corpus\ significantly reduce the noise, but there are still edge cases we cannot handle properly (e.g., the entire box glyph in a box-and-whiskers plot drawn using a single \texttt{<path>} element). The semantic labels for the same type of charts can also vary. For instance, in some bullet charts, the rectangles in the same glyph are overlapping and aligned to one side (e.g., left or bottom), while others stack the rectangles without overlapping (Figure~\ref{fig:mystique_errors}(b)). We label the former as a group with a glyph layout, and the latter as a group with a stack layout. The implications of these cases on downstream applications remain to be explored and better understood.
% \chen{some pie charts have special layout where pies are not uniformly stacked}

% We are also interested in exploring the design space of chart labels, i.e., investigating existing frameworks for describing chart representations using semantic components and developing a unified descriptive language that accommodates the diversity of general SVG charts. 
% In \corpus{}, some box-and-whisker plots draw the entire box glyphs using \texttt{<path>} elements, while others use compositions of \texttt{<rect>} and \texttt{<line>} elements. Some bullet charts align the marks in one bullet glyph to one side (e.g., left or bottom), while others stack the marks together (Figure~\ref{fig:mystique_errors}(b)). Additionally, some circle packing plots have just one layer of marks, while others have nested layers of circles overlaid.
% How to achieve a component-based label framework that is capable of reflecting both the same-type features and detailed semantic differences remains an open research question.

\parHeading{\revision{Strengthening Data Component Labeling.}} 
\revision{Currently VisAnatomy has two kinds of labels recording mappings between data and visuals: 1) \encInfo~record which visual element (e.g., mark, collection) and which channels of that element (e.g., color, position) encode data, and 2) \refEle\ record the type of data attribute (e.g., number, string) and the channel (e.g., width, x position) for each axis. For mappings that do not have associated axes or legends, information on the type of data attribute is missing and needs to be included.}

In its current state, only 392 charts (out of 942) in \corpus\ have the underlying data tables available.
% as the table is not always available from the source. 
% \leo{modify to reflect the downloaded data tables}
% especially when the chart doesn't come from a charting tool gallery. 
However, data tables offer important information such as data schema and attribute values
% ~(e.g., the number of an type of attributes) 
that can be useful for  applications like visualization redesign and 
% have been used in previous research to perform 
recommendation~\cite{viznet}. In future work, we plan to augment charts that lack underlying data tables
% with associated data tables in the following ways. 
% If the chart source contains the data table, we can directly download it and add it to \corpus.
% If the data table is not available, we would like to 
by investigating methods to automatically reconstruct these tables using existing labels such as \refEle\ and \encInfo. If 
fully automated approaches are not possible, we plan to  extend the current 
% borrow help from existing 
data extraction methods~(e.g., ChartDetective~\cite{masson2023chartdetective}) and 
% combine related functionalities with the semantic labels in \corpus\ to 
add a \textsf{\small \textbf{Data Table}} stage in our labeling tool, allowing human-machine collaborative curation of the underlying data table.

\parHeading{Labeling Inter-Element Relationships as Constraints.} Some chart scene abstraction frameworks, such as Charticulator~\cite{ren_charticulator_2018} and MSC~\cite{liu2024msc} which inspired the semantic labels in~\corpus~(Section~\ref{sec:deisredlabels}), have a \texttt{constraint} component that describes the spatial relationships between elements. Examples include \texttt{align}~(e.g., customized alignments in stacked bar charts) and \texttt{affix}~(e.g., the relative positioning between a mark and its annotations). 
% two kinds of \texttt{constraint} components that are at present not included in
\corpus~currently does not have labels on such constraints. 
The main challenge is that 
manually linking every pair of elements (e.g., a bar and its text annotation) and specifying their relationships as constraints can be time-consuming and error-prone. We need automatic algorithms to recognize and predict such constraints in batch, and  
% -processing of relationships between pairs of elements. 
% the two constraints for sets of objects and human inspection, otherwise . 
% We plan to investigate such algorithms based on \corpus, and develop 
novel interaction models to support generalizable \texttt{constraint} labeling.

\parHeading{\revision{AI-Assisted Labeling.}}  \revision{A significant portion of our time on this project was dedicated to developing and refining the labeling system. Labeling one chart using the system takes 10 to 15 minutes on average right now. The labeling efficiency can be further improved with the incorporation of AI models. In Section 4.1, we have shown that current LLMs can predict the semantic roles of SVG elements with good accuracy overall, such labels can thus be automatically populated. However, we need better interfaces for humans to verify and correct mistakes, especially for axis labels and main chart marks. We also plan to investigate additional AI support (e.g., fine-tuning LLMs using \corpus) for labeling other components like grouping and encoding.}

\parHeading{\revision{Further Enhancing the Corpus.}} 
% As we discussed in Section~\ref{subsec:GNNcase}, the current corpus size is not big enough to allow effective learning with vision features; we also expect that more examples can boost the performance of GNN-based models as more SVG hierarchies can be included. To this end, we plan to continue collecting new charts in both the SVG and bitmap formats and enlarge the size of \corpus. 
Although the number of charts in \corpus is comparable to state-of-the-art chart corpora curated using manual approaches (\Cref{sec:comparison}), and the scale of fine-grained labels can support various applications (\Cref{sec:useCases}), \corpus can be enlarged further with more chart designs and intra-type variations. 
We will open-source \corpus and our labeling tool to encourage contributions from the visualization community. 
% A significant portion of our time on this project was dedicated to developing and refining the labeling system. Given its current maturity and the fact that labeling one chart now takes 10 to 15 minutes on average, we anticipate that expanding the corpus size will not be overly time-consuming.
We have also experimented using synthetic data to augment the size of \corpus. 
% This is a commonly used technique to increase the size of many existing chart corpora such as YOLaT++~\cite{dou2024hierarchical} and FigureQA~\cite{kahou2017figureqa}. 
% \leo{any of the corpora in sec 4 used this technique? name them} 
To do this, we (1) converted charts in \corpus to scene templates in Mascot.js \cite{liu2024msc}, (2) generated compatible synthetic datasets, and (3) used Mascot.js to infuse the synthetic datasets with the templates. 
% Figure~\ref{fig:augmentExample} shows example charts generated using this approach.
Example charts are included in the \supp.
We decide not to include these generated charts in \corpus: the distribution of the synthetically generated data should be determined according to the specific machine-learning task or interactive application; it is better that \corpus only contains the original real-world charts, which can be augmented in different ways depending on the use case. 
% Second, currently the implementation of the repopulate operator in Mascot.js cannot handle all the 40 chart types in \corpus yet.   
Future research may develop approaches to promoting diversity in the underlying data and the visual styles at the same time, and to achieve a desired balance between quantity and diversity. 

% Finally, we intend to investigate methods for learning-based data augmentation. 
% to avoid too much manual effort is a worthwhile endeavor.  Specifically, we would like to explore if GNN can model the chart layout specified together by \layoutInfo\ and \encInfo~as templates, and transform a given SVG chart into a new one with similar visual designs but representing a new data table. 

% In this way, we may also be able to obtain the semantic labels for the new chart automatically, avoiding the need for manual labeling. 
% We also plan to experiment with large language models~(e.g., fine-tuning LLMs using \corpus) to see if LLMs are capable of helping accelerate the labeling process by regarding the labeling task as a code generation task. 

\revision{
\corpus does not yet include node-link visualizations. The main challenge in labeling is similar to that in labeling constraints: automatic algorithms are needed to recognize and predict relationships between nodes and links. Composite visualizations, where one chart overlays on another with each chart having its own axis, are not within the scope either. By focusing on SVG charts, we may also have omitted unique visualization designs that are available only in formats like raster images. \corpus can be further enriched by including these missing charts, with AI-assisted methods to label components.}

\parHeading{Enhancing the Breadth and Depth of Applications.} 
% We plan to explore more visualization applications in greater breadth and depth to unlock the full capacity of \corpus. 
The GNN models tested in~Section~\ref{subsec:GNNcase} are homogeneous, having the same feature length for all the nodes. 
% However, given that SVGs are composed of a variety of different elements that have distinct sets of visual properties, heterogeneous GNNs~\cite{HeGNN}, which accept 
Supporting node features of unequal lengths for different types of nodes could be a better approach. Second, the semantic labels have the potential to support the development of rigorous approaches to computing pair-wise chart similarities, which can serve as a quantitative measure to evaluate corpus diversity~\cite{chen2023state}, guiding the future collection of chart corpora. Third, for research that combines Visualization and Natural Language ~(NL)  to solve tasks such as chart QA~\cite{kahou2017figureqa} and chart captioning~\cite{mahinpei2022linecap},  the diverse set of charts in \corpus\ and the associated semantic labels can be utilized to synthesize datasets that contain high-quality (VIS, NL) pairs,
% adequate linguistic variations for a large variety of charts, 
enhancing the generalizability of multi-modal models.
% to real-world human-asked questions.

\parHeading{Supporting Interaction and Animation Labeling.} Currently \corpus focuses on semantic labels in static charts. A future direction is to annotate interaction and animation. The semantic labels of static components are based on existing visualization abstraction models; we envision that the annotation of interaction and animation also requires solid theoretical foundations on dynamic visualizations. We will investigate how interaction grammars like Vega-Lite \cite{satyanarayan_vega-lite_2016} may be used and explore additional abstractions. 

% \subsection{Limitations}
% corresponding data table can be hard to get, ChartDetective

% augmentation: we release the labeling tool; LLM-assisted crowdsourcing is possible while is performing bad since it fakes urls 

% there are two kinds of svg representations for a bullet chart glyph: 1) rects with different widths aligned to the left, 2) rects with different widths stacked. In the second case, we annotated both width and left as the channels in encodings.

% Box and whisker plot: sometimes the entire glyph is drawn as a path, difficult/impossible to annotate

% layout: current layout type cannot capture the rich variations in layout, e.g., circle packing example, word cloud example

% \subsection{Complete Annotations}
% relational constraints: fixation, alignment, ... 

% the design space of chart annotations

% \subsection{Increase the Corpus Size}
% continuing curation + annotation, open source to get contributions from the community

% data augmentation techniques

% \subsection{Explore additional use cases/applications}

% \chen{in discussion section can talk about Heterogeneous GNNs}

% use LLM to generate QA datasets

% chart simnilarity

% \mytodo{}
\section{Conclusion}

% Most existing chart corpora in the visualization community have limitations that hinder their ability to support a wide range of downstream applications. These limitations stem from two primary aspects: (1) the restricted set of chart types included and (2) the lack of fine-grained semantic labels. 
In the paper, we contribute \corpus, a diverse SVG chart corpus encompassing 40 chart types produced by over 50 tools from hundreds of public online sources. Each chart in \corpus\ is augmented with rich, multi-granular semantic labels including graphical elements' types, roles, and bounding boxes, the hierarchical grouping of elements, the layouts of groups, and visual encodings. 
We compare \corpus\ with related corpora to demonstrate its diversity and the richness of the semantic labels. The usefulness of \corpus\ is evaluated through four applications: \revision{semantic inference of SVG element roles}, chart semantic decomposition, chart type classification, and content navigation for accessibility. Finally, we outline research challenges and opportunities for future work.
\acknowledgments{%
  This work was supported by NSF grant
IIS-2239130.%
}

\bibliographystyle{abbrv-doi-hyperref}

\bibliography{_main}

\appendix % You can use the `hideappendix` class option to skip everything after \appendix

% \section{About Appendices}
% Refer to \cref{sec:appendices_inst} for instructions regarding appendices.

% \section{Troubleshooting}
% \label{appendix:troubleshooting}

% \subsection{ifpdf error}

% If you receive compilation errors along the lines of \texttt{Package ifpdf Error: Name clash, \textbackslash ifpdf is already defined} then please add a new line \verb|\let\ifpdf\relax| right after the \verb|\documentclass[journal]{vgtc}| call.
% Note that your error is due to packages you use that define \verb|\ifpdf| which is obsolete (the result is that \verb|\ifpdf| is defined twice); these packages should be changed to use \verb|ifpdf| package instead.

% \subsection{\texttt{pdfendlink} error}

% Occasionally (for some \LaTeX\ distributions) this hyper-linked bib\TeX\ style may lead to \textbf{compilation errors} (\texttt{pdfendlink ended up in different nesting level ...}) if a reference entry is broken across two pages (due to a bug in \verb|hyperref|).
% In this case, make sure you have the latest version of the \verb|hyperref| package (i.e.\ update your \LaTeX\ installation/packages) or, alternatively, revert back to \verb|\bibliographystyle{abbrv-doi}| (at the expense of removing hyperlinks from the bibliography) and try \verb|\bibliographystyle{abbrv-doi-hyperref}| again after some more editing.

\end{document}